# A Sampling Strategy Benchmark for Machine-Learning-Based Seismic Liquefaction Prediction

Jilei Hu[a, b, c, *], Fenglin He[b], Lianming Huang[b], Qianfeng Wang[b]

[a] Key Laboratory of Geological Hazards on Three Gorges Reservoir Area, Ministry of Education, Three Gorges University, Yichang 443002, Hubei, China

[b] College of Civil Engineering and Architecture, China Three Gorges University, Yichang, Hubei 443002, China

[c] Key Laboratory of Soft Soils and Geoenvironmental Engineering (Zhejiang University), Ministry of Education

**Abstract:** Sampling strategy including sampling methods and training set configurations (training set sample size, train-test split ratio, and class distribution) significantly affects machine-learning (ML) model performance in seismic liquefaction prediction. However, existing ML applications in seismic liquefaction prediction remain fragmented: sampling strategies vary widely across studies without a unified benchmark. Moreover, these studies generally optimize the sample set configuration independently, ignoring the interaction among training set configurations. To address these limitations, this study establishes a benchmark that systematically evaluates sampling methods, training set sample sizes, train-test split ratios, class distributions, and training set configurations coupling on seven mainstream ML models performance, and further improves the predictive accuracy of seismic liquefaction — using a database of 250 historical liquefaction events, evaluated by $Acc$ and $F_1$. The results show that ordered systematic sampling yields the best performance across all models. The optimal model can be trained when the training set sample size is 200, the train-test split ratio is 80:20, and the class distribution range is 1~1.5. Among them, the train-test split ratio most significantly influenced performance, followed by the class distribution, with the training set sample size having the least effect. Furthermore, the Random Forest model achieves the highest performance, while the K-Nearest Neighbor model performs the weakest. Importantly, this study systematically identifies and verifies for the first time that there will be an interaction effect among training set configurations, rather than a simple additive effect. This study provides a benchmark for scholars to select the optimal sampling method and training set configurations to obtain high accuracy in ML-based liquefaction prediction.

---

[*] Corresponding author's E-mail: hujl@ctgu.edu.cn (J.L. Hu)

1. **Introduction**

Earthquake-induced liquefaction usually brought major disasters to buildings and infrastructures during several historical earthquakes [1]. Therefore, the accuracy of liquefaction prediction is critical. Machine learning strategies can be utilized to analyze extensive datasets, uncovering underlying correlations related to seismic sand liquefaction [2]. This approach enables accurate identification and prediction of liquefaction-prone areas, thereby enhancing prediction accuracy.

Up to now, all studies have been trying to improve the prediction accuracy of the liquefaction models by using different sampling strategies and ML methods such as logistical regression [3], artificial neural network [4], support vector machine [5], Bayesian network [6], random forest [7], etc., and studying the selection of feature variables[1], [8]. Table 1 summarizes the relevant literature on ML methods and sampling strategies in seismic liquefaction prediction. However, these studies ignored the effect of different sampling strategies on the performance of probabilistic liquefaction models. For instance, Nejad et al. [9] and Zhou et al. [10] used a stratified sampling method to divide the dataset. Demir and Şahin [11] used simple random sampling, stratified sampling, and oversampling methods to divide the dataset and pointed out that different sampling methods will have an impact on the model performance. They did not introduce the details of what sampling strategies were used in their model training except for Hu and Liu [8] who mentioned the stratified sampling technology for guaranteeing the distribution of training data same as the distribution of all data. Unfortunately, they did not discuss the effect of different sampling strategies on the performance of models.

Table 1. Comparison of the sampling strategies in the references.

| References | Models and algorithms | Sampling methods | Sample sizes | Train-test split ratios | Class distributions |
|---|---|---|---|---|---|
| Hu et al. [17] | BN | Stratified sampling | 125 | 1:4 | 1.5:1 |
| | | | 150 | 1:2 | |
| | | | 200 | 1:1 | |
| | | | 250 | 1.5:1 | |
| | | | 300 | 2:1 | |
| | | | 350 | 2.5:1 | |
| | | | 400 | 3:1 | |
| | | | 450 | 3.5:1 | |
| | | | 500 | 4:1 | |
| Zhou et al. [10] | RF | Stratified sampling | 226 | 80:20 | 133:93 |
| | | | 620 | | 256:364 |
| | | | 415 | | 287:128 |

Continued Table 1. Comparison of the sampling strategies in the references.

| References | Models and algorithms | Sampling methods | Sample sizes | Train-test split ratios | Class distributions |
|---|---|---|---|---|---|
| Bherde et al. [21] | RF, DT, and LR | Random sampling | 225 | 70:30 | / |
| | | | 188 | 75:25 | |
| | | | 159 | 80:20 | |
| Chithuloori and Kim [22] | RF | / | 540 | 70:30 | 1:1 |
| Demir and Şahin [16] | RF | Stratified random sampling | 226 | 50:50 | 133:93 |
| | | | | 60:40 | |
| | | | | 70:30 | |
| | | | | 50:50 | |
| | | | 253 | 60:40 | 181:72 |
| | | | | 70:30 | |
| Ghani et al. [24] | RF and DT | / | 210 | 80:20 | 113:97 |
| Nguyen et al. [15] | ANN and DT | Random sampling | 538 | 10:90 | / |
| | | | | 20:80 | |
| | | | | 30:70 | |
| | | | | 40:60 | |
| | | | | 50:50 | |
| | | | | 60:40 | |
| | | | | 70:30 | |
| | | | | 80:20 | |
| | | | | 90:10 | |
| Ahmad M et al. [18] | BN and DT | Random sampling | 50 | 80:20 | 2.5:1 |

Note: slashes represent unknown information in some references; class distribution represents the ratio of liquefied samples to non-liquefied samples.

In addition to the sampling strategies, training set configurations (training set sample size, train-test split ratio, and class distribution) also have an impact on the model performance. Previous studies have demonstrated that smaller sample sizes may result in overfitting or underfitting, while larger sample sizes can enhance the model's robustness and generalization [10], [12][14]. In the existing research on seismic liquefaction discrimination, scholars also compared the effect of different train-test split ratios on the performance of models, and the results showed that the performance of models showed variability with different train-test split ratios [15][20]. Moreover, historical datasets typically exhibit class imbalance. While Oommen et al. [26] demonstrated that optimal LR performance occurs when sample class distributions match overall distributions using synthetic data, this finding requires validation for actual seismic field data. Jain [27] extended the aforementioned information to real case histories of soil liquefaction by sampling seven different distributions in the two datasets (population distributions of liquefaction to nonliquefaction are similar to 1:1 for SPT data and 3:1 for CPT data): 20:80 (liquefaction: nonliquefaction), 30:70, 40:60, 50:50, 60:40, 70:30, and 80:20. The highest $P$ value (0.99) obtained for a sample distribution of 50:50 (1:1) is closest to the population distribution of 1:1 for the SPT data set, which verified the conclusion of Oommen et al. [26]. However, the highest $P$ value (0.99) obtained for a sample distribution of 40:60 (1:1.5) is quite different from the population distribution of 3:1 for the CPT data set, which went against the conclusion of Oommen et al. [26]. Hu et al. [28] investigated 11 different distributions on the performance of four ML models, and found that the optimal class distribution is not a fixed value but a range.

The above studies show that the field is still fragmented, and a unified benchmark that can jointly evaluate the sampling method and training set configurations is urgently needed. In addition, previous studies generally optimize the sample set configuration in isolation, ignoring the interaction among the training set configurations. Therefore, establishing a benchmark to quantitatively evaluate sampling methods and training set configurations is essential for advancing machine-learning-based seismic liquefaction prediction.

The study aims to investigate: (1) the impact of different sampling methods on the performance of ML models, and to determine the optimal sampling method; (2) the impact of different training set sample sizes on the performance of ML models, and to determine the optimal training set sample size; (3) the impact of different train-test split ratios on the performance of ML models, and to determine the optimal train-test split ratio; (4) the impact of different class distributions on the performance of ML models, and determine the optimal class distribution; (5) the effects of training set configurations coupling on the performance of ML models; (6) a comprehensive benchmark comparison of 7 mainstream ML models within the framework of seismic liquefaction prediction. The results of the study can provide a scientific reference basis for scholars to choose the optimal sampling method and training set configurations to improve the accuracy of liquefaction prediction, while contributing a rigorous multi-model benchmark for ML-based liquefaction prediction.

## 2. Methodologies

2.1 Sampling strategies

Sampling methods mainly fall into probability and non-probability categories. Considering the subjectivity of non-probability sampling, the following five probability sampling methods are selected for this study: simple random, unordered system, ordered system, stratified random and whole cluster sampling, all detailed in Table 2.

Table 2. Probability sampling methods.

| Sampling method | Theory |
|---|---|
| Simple Random Sampling (SRS) | SRS involves selecting individuals from the population with equal and independent probability. In this study, five subsets of 50 samples each were generated through simple random sampling; one subset is used as the test set each time, and the remaining four as the training set. |
| Unordered Systematic Sampling (USS) | USS ignores the order in which the totals are arranged when selecting samples, and samples are selected at fixed intervals after the factors in the totals that are relevant to the study have been randomly disrupted. In this study, time was chosen as the basis for ordering. |
| Ordered Systematic Sampling (OSS) | OSS is to sort all the influencing factors according to their earthquake impact weights, starting with liquefaction status, followed by $q_{c1Ncs}$, $PGA$, $\sigma'$, $FC$, $D_s$, $M_w$, and $D_w$. The sorted dataset is divided into five subsets at fixed intervals of 5; each subset is used once as the test set, with the remaining four as the training set. |
| Stratified Random Sampling (StrRS) | StrRS divides the population into strata based on specific criteria, then samples each stratum proportionally. Here, stratification was based on $q_{c1Ncs}$, and five non-overlapping subsets were formed through proportional random sampling. |
| Cluster Sampling (CS) | CS is the process of grouping the units in the population into subgroups that do not cross each other or repeat each other, and then sampling the units as clusters. In this study, the data were classified into different clusters according to regions. |

2.2 Machine learning models

2.2.1 Logistical Regression

Logistic Regression (LR) is one of the most commonly used statistical methods in research. The algorithm incorporates a linear regression model into a logistic function, commonly used as a sigmoid function, which has the following form (Eq. 1):

$$y = \frac{1}{(1+\exp^{-x})} \quad (1)$$

Figure 1 depicts the Sigmoid function's S-shaped curve exhibiting asymptotic behavior at 0 and 1. For liquefaction probability $P_L$ under $n$ influencing factors $X = (x_1, x_2, …, x_n)$, the multifactorial probability model is given by Eq. 2:

$$P_L(X) = P_L[Y=1 \mid X] = E[Y \mid X], 0 \leq P_L \leq 1 \tag{2}$$

The complex non-linear relationship between the influencing factors and the predicted probability of liquefaction is transformed into a linear relationship between the independent variables and logit, which can be represented by a binomial logistic regression model, with the following formula (Eq. 3):

$$\log it(P_L) = \ln(\frac{P_L}{1-P_L}) = \beta_0 + \beta_1 x_1 + \beta_2 x_2 + \cdots + \beta_n x_n \tag{3}$$

where $P_L$ is the probability that liquefaction of the sandy soil occurs; $x_1, x_2, \ldots, x_n$ are the independent variables; $\beta_0, \beta_1, \beta_2, \ldots, \beta_n$ are the constant terms and the coefficients of the respective variables, respectively. An exponential transformation of Eq. 3 yields an expressed relationship between the independent variables and the results of the probability calculations, with the following formula (Eq. 4):

$$P_L = \frac{1}{1+\exp[-(\beta_0 + \beta_1 x_1 + \beta_2 x_2 + \cdots + \beta_n x_n)]} \tag{4}$$

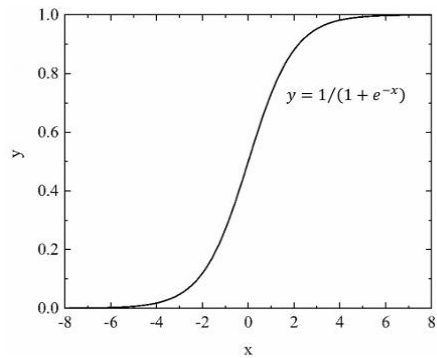

Figure 1. Sigmoid function curve schematic.

2.2.2 Artificial Neural Network

Artificial Neural Networks (ANN) are based on data and information-driven computational models. The basic parts that make up an artificial neural network are called artificial neurons, also known as perceptual machines. Table 2 shows the classical artificial neural network structure, which consists of an input layer, an output layer, and multiple hidden layers. The training of the neural network is divided into the following steps:

(1) Non-linear transformation of the input data layer by layer through the activation function and output of the prediction results, calculated by the following formula (Eq. 5):

$$v_j^{(i)}(n) = \sum_{i=0}^{T} w_{ji}^{(i)}(n) y_i^{i-1}(n) \tag{5}$$

where $y_i^{i-1}(n)$ is the work signal transmitted from unit $i$; T is the number of inputs before adding to unit $j$; $w_{ji}^{(i)}(n)$ is the connection right between node $i$ and node $j$.

(2) The computational error is obtained by subtracting the prediction result obtained from the computation from the true value, back-propagating the error to the input layers, and calculating the gradient at each layer, calculated by the following formula (Eq. 6):

$$\delta_j^{(i)}(n) = e_i^{(L)}(n)O_j(n)[1-O_j(n)] \tag{6}$$

where $\delta_j^{(i)}(n)$ is the local gradient; $O_j(n) = y_j^{(0)}(n)$.

(3) Based on the calculated gradient, the weights are updated using an algorithm to obtain the minimized error function, calculated by the following formula (Eq. 7):

$$w_{jk}^{(i)}(n+1) = w_{jk}^{(i)}(n) + \eta\delta_j^{(i)}(n)y_i^{i-1}(n) \tag{7}$$

where $\eta$ is the learning step size and $n$ is the number of training sessions. Repeat the above steps until the stopping criterion is met.

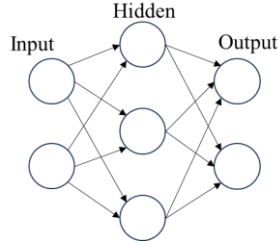

Figure 2. Schematic diagram of an artificial neural network structure.

2.2.3 Support Vector Machine

Support Vector Machine (SVM) is a simple binary classification model, which is shown in Figure 3. Its basic structure is defined as an interval-maximizing linear classifier on the feature space, which transforms the input space to a high-dimensional space using a nonlinear transformation and finds an optimal classification surface in that space with interval maximization as the principle. Since the vast majority of practical engineering problems are nonlinear, the estimation function is given by the following formula (Eq. 8):

$$f(x) = W \cdot \Phi(x) + b \tag{8}$$

where $W$ is the adjustable weight function, $b$ is the function bias value, $\Phi(x)$ is a nonlinear mapping function, and both are $n$-dimensional vectors with $W$. To find $W$ and $b$, the relaxation factors $\xi_1$ and $\xi_2$ are introduced to obtain Eq. 9.

$$\min Q = \frac{1}{2}\|\omega\|^2 + C\sum_{i=1}^{n}(\xi_1 + \xi_2) \tag{9}$$

where $\|\omega\|^2$ is the descriptive function; $C$ is the penalty factor.

Combining the KKT condition and the dyadic condition, the original function is optimized to give the expression (Eq. 10):

$$f(x) = \sum_{i,j=1}^{n}(\alpha_i - \alpha_j^*)k(x_i, x_j) + b \tag{10}$$

where $\alpha_i$ and $\alpha_j^*$ are lagrange factors; $k(x_i, x_j)$ is the radial basis function.

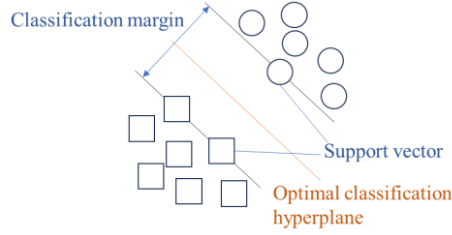

Figure 3. Support vector machine model diagram.

2.2.4 Bayesian Network

Bayesian Network (BN) is a directed acyclic probabilistic graphical model based on causal inference. A Bayesian network consists of three components: nodes, directed lines, and probability tables, where nodes represent arbitrary variables, directed lines represent causal dependencies between nodes, and probability tables quantify the probabilistic relationships between connected nodes. Let there exist a directed acyclic graph $G$ that can represent the relationship between variables $X_i$ ($1 \leq i \leq n$), then the joint probability of node $X_i$ can be expressed as Eq. (11):

$$p(x) = \prod_{i \in n} p(X_i | pa(X_i)) \tag{11}$$

where: $pa(X_i)$ denotes the factor of node $X_i$ or the set of parent nodes. For any random variable, the joint probability can be expressed as Eq. (12):

$$p(X_1, \cdots, X_i) = p(X_i | X_1, \cdots, X_{i-1}) \cdots p(X_2 | X_1)p(X_1) \tag{12}$$

The joint probability distribution of each node can be derived when the network structure and parameters in a Bayesian network are known, and a simple BN model is shown in Figure 4.

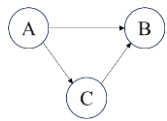

Figure 4. A simple Bayesian network model diagram.

2.2.5 Decision Trees

Decision Trees (DT) can flexibly handle a variety of data types, including categorical and continuous data, and they classify or regress data through the judgment of a series of characteristic attributes. Figure 5 shows the schematic diagram of a single decision tree binary classification model, the topmost in the figure is the bottom of the root node, which contains all the sample data, and the data is divided into the middle level of child nodes according to a certain attribute of that root node; and so on, from top to bottom, to divide the data into the category to which it belongs. Therefore, the key to constructing a decision tree is to select the appropriate attribute in the current state as the node to divide the data into categories, and to reach the leaf nodes in the fastest way according to a certain objective function (e.g., information entropy, Gini coefficient, etc.), to make a final judgment on the data categories.

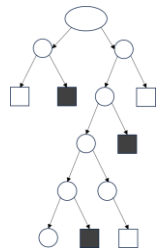

Figure 5. Decision Tree model diagram.

2.2.6 Random Forest

Random Forest (RF) is an ensemble algorithm combining improved decision trees with randomized feature selection. This design enables robust multi-factor classification. As shown in Figure 6, RF implementation involves two randomized processes:

(1) When collecting data for training from the sample data, $k$ new sets of training samples are randomly selected by the Bootstrap method, and $k$ decision trees are constructed from them.

(2) During tree construction, RF randomly selects a feature subset at each node for splitting, not all features. This reduces model variance and increases tree diversity. Dual randomization enhances forest diversity through amplified inter-tree variance.

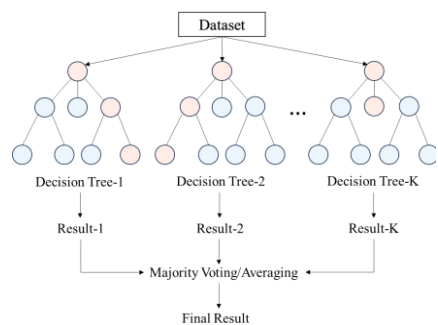

Figure 6. Random Forest model diagram.

2.2.7 K-Nearest Neighbor

K-Nearest Neighbor (KNN) is a non-parametric classification method employing lazy learning. It classifies test instances by majority voting among the $k$ closest training samples, constructing decision boundaries ad-hoc without explicit model training. Suppose the training set $X = \{x_1, x_2, \ldots, x_n\}$ is a set of points in a multidimensional space with the classification attribute $l$. It is a discrete variable with a value domain $L = \{1, 2, \ldots, k\}$, and the goal of classification is to minimize the classification error $M_j$, calculated by the following formula (Eq. 13):

$$M_j = \sum_{l_j \in L} R_{l_j l_j'} p(l_j' \mid q) \tag{13}$$

where $R_{l_j l_j'}$ represents the error caused by classifying the value $l_j$ as $l_j'$ ($j \neq j'$) and $q$ represents the predicted point $p(l_j' \mid q)$ represents the probability of classifying q as $l_j'$. In general, KNN assumes that all misclassifications have the same error, with the following formula (Eq. 14):

$$R_{l_j l_j'} = \begin{cases} 0, j = j' \\ 1, j \neq j' \end{cases} \tag{14}$$

Whereas the KNN method does not accurately predict the value of the classification attribute of $q$, but gives the most likely prediction, with the following formula (Eq. 15):

$$\text{KNN}(q) = \arg\max_{l_j \in L} p(l_j \mid q) \tag{15}$$

where KNN($q$) denotes the KNN prediction for point $q$, the classification procedure is: (1) Select distance metric; (2) Identify $k$ nearest training neighbors for each test point $s$; (3) Assign $q$'s class via majority voting among neighbors. Classification accuracy is then evaluated using Eq. 13.

2.3 *K*-fold cross-validation

*K*-fold cross-validation mitigates performance metric variance induced by random training-test partitioning, thus providing unbiased model evaluation. Figure 7 shows the process of *K*-fold cross-validation, which is divided into the following four main steps: (1) The database is divided into *K* parts by sampling, each having the same sample size. (2) *K*-1 fold is used to train the model, and the remaining fold is used to validate the performance of the model. The process is repeated *K*-1 times so that all the samples in the database are used for training and validation. (3) After *K* trials, the average of all trial results is used as the final prediction of the model.

When *K* is taken as 5, the training set has reached 80%, which is close to the performance of complete data training. When *K* is taken as 10, the sample size of the test set is only 25, which may lead to unstable results, and the larger the value of $K$, the lower the variance of the evaluation results will be. $K = 5$ can provide low bias and controllable variance of the performance evaluation with low computational cost while avoiding the instability of the small dataset. Therefore, 5-fold cross-validation was chosen in this study.

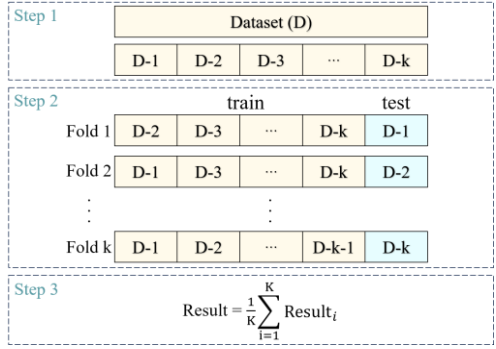

Figure 7. K-fold cross-validation method.

2.4 Information entropy

Information entropy (*IE*) is a measure of the purity or uncertainty of a dataset, reflecting the average uncertainty of the information in the dataset. In classification problems, information entropy is used to quantify the uniformity of the distribution of categories in a dataset (Eq. 16):

$$\mathrm{IE} = -\sum_{i=1}^{n} p_i \log_2 p_i \tag{16}$$

where $p_i$ is the probability of a sample belonging to the $i$th category in the dataset, and $n$ is the total number of categories. The lower the information entropy, the higher the purity of the dataset; conversely, the higher the information entropy, the higher the uncertainty of the data.

Joint entropy $H(X)$ is mainly used to examine the content of information shared among multiple variables, for a joint variable consisting of multiple random variables $(X_1, \ldots, X_n)$. Their joint entropy is expressed by Eq. 17 as follows:

$$H(X_1, \cdots, X_n) = -\int_{x_1} \cdots \int_{x_n} p(x_1, \cdots, x_n) \log p(x_1, \cdots, x_n) dx_1 \cdots dx_n \tag{17}$$

2.5 Kullback-Leibler

Kullback-Leibler (KL) divergence can also be referred to as relative entropy. The theoretical significance of the KL divergence is to measure the degree of difference between two probability distributions. The KL divergence value can be expressed as Eq. 18:

$$\mathrm{KL}(P \parallel Q) = \int p(x) \log \frac{p(x)}{q(x)} dx \tag{18}$$

where *x* is a random variable; *p(x)* and *q(x)* are two probability density functions of the same variable in the two data sets; *P* and *Q* are the corresponding probability distributions of *p(x)* and *q(x)*. The greater the KL divergence value, the greater the degree of difference between the two, and the lesser the KL divergence value, the greater the similarity between the two.

2.6 Performance measures

This study is a binary classification problem, and the results can be represented by a confusion matrix. The confusion matrix consists of four components: True Positive (TP): correct predictions of liquefied soil; True Negative (TN): correct predictions of non-liquefied soil; False Positive (FP): incorrect predictions of liquefied soil; and False Negative (FN): incorrect predictions of non-liquefied soil. Common performance evaluation metrics in classification models are accuracy, precision, recall, and $F_1$-score; the formulae are as follows.

Accuracy is the percentage of samples that are correctly predicted out of all samples (Eq. 19):

$$Acc = \frac{TP + TN}{TP + TN + FP + FN} \quad (19)$$

The precision rate is the probability of a positive sample among all predicted positives, i.e., the certainty of correct predictions among positive predictions, defined by Eq. 20:

$$Pre = \frac{TP + TN}{TP + TN + FP + FN} \quad (20)$$

Recall is the probability of being predicted as positive in a sample that is positive (Eq. 21):

$$Rec = \frac{TP + TN}{TP + TN + FP + FN} \quad (21)$$

*Pre* and *Rec* are sometimes mutually exclusive, i.e., when *Pre* is high, *Rec* falls. To balance *Pre* and *Rec*, the most common method is *F*-Score, which is a weighted average of *Pre* and *Rec* (Eq. 22):

$$F_1 = \frac{2 * Pre * Rec}{Pre + Rec} \quad (22)$$

Because $F_1$ integrates *Pre* and *Rec*, *Acc* and $F_1$ are selected as the performance evaluation indexes of the model in this study.

## 3. Data collection, experimental design, and model construction

3.1 Liquefaction data sources and collection

In this study, 250 CPT historical seismic liquefaction datasets were collected, consisting of 178 liquefaction datasets and 72 non-liquefaction datasets [29]. These data correspond to 13 earthquakes that occurred in the period from 1964 to 2011 and include seven regions: Japan, California, the United States, the United Kingdom, Turkey, Taiwan, Xinjiang, China, and New Zealand. The dataset contains seven influencing factors: earthquake magnitude ($M_w$), peak ground acceleration ($PGA$), fine-grain content ($FC$), overlying effective stress ($\sigma'$), depth to water table ($D_w$), depth of burial of sandy layer ($D_s$), and equivalent clean sand cone-head resistance ($q_{c1Ncs}$). The range of these data is described in Table 3. Figure 8 shows a heat map of the correlations between the variables, with correlation coefficients no greater than 0.8 for each pair of variables, with only $D_s$ and $\sigma'$ strongly correlated, $D_w$ and $\sigma'$ moderately correlated, and the rest of the variables are only weakly or very weakly correlated. Therefore, the variables used in this study are $M_w$, $PGA$, $FC$, $\sigma'$, $D_w$, $D_s$, and $q_{c1Ncs}$.

Table 3. Description of the data.

| Parameter | Notation | Unit | Min | Max |
|---|---|---|---|---|
| Earthquake magnitude | $M_w$ | - | 5.9 | 7.6 |
| Peak horizontal Acceleration | $PGA$ | g | 0.08 | 0.80 |
| Fines content | $FC$ | % | 0 | 98.4 |
| Effective vertical stress | $\sigma'$ | kPa | 17.3 | 149.4 |
| Groundwater table depth | $D_w$ | m | 0.4 | 6.4 |
| The depth of the soil layer | $D_s$ | m | 1.5 | 11.0 |
| The equivalent corrected tip resistance | $q_{c1Ncs}$ | - | 21.0 | 254.0 |

3.2 Experimental design

3.2.1 Experimental design for the effect of sampling strategies on model performance

Five probabilistic sampling methods (SRS, USS, OSS, StrRS, and CS) partitioned the 250-case dataset into five 50-sample folds. Using standard 5-fold cross-validation (Figure 8), four-fold trained models, while one-fold tested, were repeated five times. Model performance was evaluated via averaged $Acc$ and $F_1$ across iterations to determine the optimal sampling method.

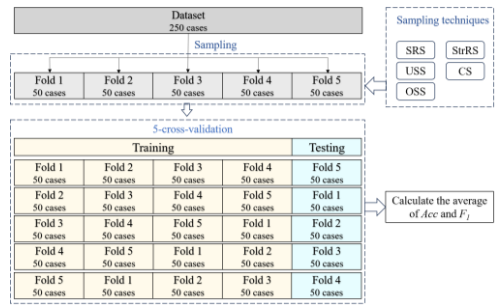

Figure 8. Schematic design of experiments on the effect of sampling methods on model performance.

3.2.2 Experimental design for the effect of training set sample size on model performance

To isolate the impact of training set sample size on model performance while maintaining a consistent train-test split ratio, a fixed ratio of 4:1 was adopted throughout this experiment, thereby leveraging the inherent dataset partitioning structure of the 5-fold cross-validation methodology employed in previous steps. To study the effect of the total sample size of 50, 100, 150, 200, and 250 (i.e., the training set sample size of 40, 80, 120, 160, and 200) on the model performance under the optimal sampling method, the experimental design is shown in Figure 9.

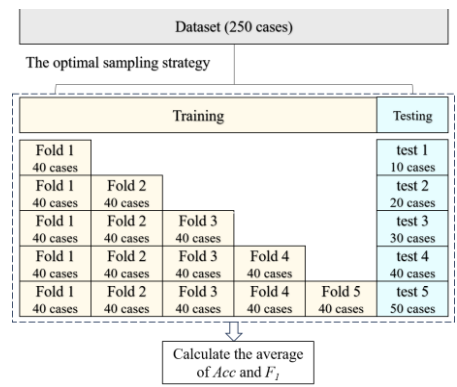

Figure 9. Schematic design of experiments on the effect of training set sample size on model performance.

3.2.3 Experimental design for the effect of train-test split ratio on model performance

In previous studies, common training-to-test set ratios were 60:40, 70:30, and 80:20, and in addition to these three ratios, 50:50 and 90:10 were also investigated in this paper. According to five different training-testing set ratios (50:50, 60:40, 70:30, 80:20, and 90:10) under the optimal sampling method, 250 samples are divided into training and testing sets, and the experimental design is shown in Figure 11 to study the effect of different training-testing set ratios on the model performance.

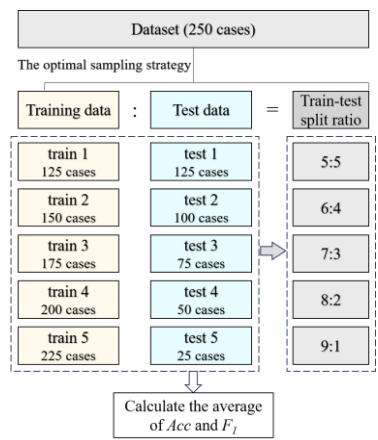

Figure 10. Schematic design of experiments on the effect of train-test split ratios on model performance.

3.2.4 Experimental design for the effect of class unbalance on model performance

Hu et al. [28] investigated the performance of ML models with 11 different class distributions (10:90, 20:80, 25:75, 33:67, 40:60, 50:50, 60:40, 67:33, 80:20, and 90:10). Its study shows that the performance of the model is significantly poor when the class distribution is large (90:10 and 10:90). Therefore, this study builds on this study to investigate the model performance with nine different class distributions: 20:80 (1:4), 25:75 (1:3), 33:67 (1:2), 40:60 (1:1.5), 50:50 (1:1), 60:40 (1.5:1), 67:33 (2:1), 75:25 (3:1) and 80:20 (4:1) on model performance. Since the ratio of liquefied to non-liquefied samples in the available dataset (2.5:1) was not satisfactory for the study of class imbalance, the sample size was kept at 250, and the ratio of liquefied to non-liquefied samples in it was adjusted to (1.25:1). Based on the optimal sampling method, to study only the effect of class imbalance on model performance in the training set, the class ratio of the test set was kept as 1:1. The experimental design is shown in Table 4.

Table 4. Schematic design of experiments on the effect of class unbalance on model performance.

| | Training | | Testing | |
|---|---|---|---|---|
| Class distribution | Data size (Lip.) | Data size (Non-Lip.) | Data size (Lip.) | Data size (Non-Lip.) |
| 1:4 | 40 | 160 | 25 | 25 |
| 1:3 | 50 | 150 | 25 | 25 |
| 1:2 | 66 | 134 | 25 | 25 |
| 1:1.5 | 80 | 120 | 25 | 25 |
| 1:1 | 100 | 100 | 25 | 25 |
| 1.5:1 | 120 | 80 | 25 | 25 |
| 2:1 | 134 | 66 | 25 | 25 |
| 3:1 | 150 | 50 | 25 | 25 |
| 4:1 | 160 | 40 | 25 | 25 |

Note: Liq. and Non-Liq. mean liquefaction and non-liquefaction, respectively.

3.2.5 Experimental design for the effect of training set configurations coupling on model performance

In practice, the training set sample size, train-test split ratio, and class distribution may be changing simultaneously. The effect of these training set configurations coupling on the model performance is not clear, so based on the optimal sampling method, three training set sample sizes, three train-test split ratios, and three class distributions with better model performance are selected for the coupling study, and there are 27 different coupling cases, as shown in Table 5.

Table 5. Design of experiments on the effect of training set configurations coupling on model performance.

| Cases | Training set sample size | Train-test split ratio | Class distribution |
| --- | --- | --- | --- |
| 1 | 120 | 60:40 | 50:50 |
| 2 | 120 | 60:40 | 60:40 |
| 3 | 120 | 60:40 | 67:33 |
| 4 | 120 | 70:30 | 50:50 |
| 5 | 120 | 70:30 | 60:40 |
| 6 | 120 | 70:30 | 67:33 |
| 7 | 120 | 80:20 | 50:50 |
| 8 | 120 | 80:20 | 60:40 |
| 9 | 120 | 80:20 | 67:33 |
| 10 | 160 | 60:40 | 50:50 |
| 11 | 160 | 60:40 | 60:40 |
| 12 | 160 | 60:40 | 67:33 |
| 13 | 160 | 70:30 | 50:50 |
| 14 | 160 | 70:30 | 60:40 |
| 15 | 160 | 70:30 | 67:33 |
| 16 | 160 | 80:20 | 50:50 |
| 17 | 160 | 80:20 | 60:40 |
| 18 | 160 | 80:20 | 67:33 |
| 19 | 200 | 60:40 | 50:50 |
| 20 | 200 | 60:40 | 60:40 |
| 21 | 200 | 60:40 | 67:33 |
| 22 | 200 | 70:30 | 50:50 |
| 23 | 200 | 70:30 | 60:40 |
| 24 | 200 | 70:30 | 67:33 |
| 25 | 200 | 80:20 | 50:50 |
| 26 | 200 | 80:20 | 60:40 |
| 27 | 200 | 80:20 | 67:33 |

3.3 Model construction

In this study, seven ML models (LR, ANN, SVM, BN, DT, RF, and KNN) are constructed in the IBM SPSS Modeler 18.0 based on 250 datasets that have been collected. The influences of the model inputs are $M_w$, $PGA$, $FC$, $\sigma'$, $D_w$, $D_s$, and $q_{c1Ncs}$. The model outputs are all about whether or not to liquefy. The construction of the models in the IBM SPSS Modeler is shown in Figure 11. The hyperparameters of each model were determined through empirical tuning, as shown in Table 6. In order to avoid the influence of hyperparameters on the model performance, it was ensured that the hyperparameters of each model were consistent.

Table 6. Model hyperparameters.

| Machine learning methods | Model | Model hyperparameters |
|---|---|---|
| Random Forest | RF | max-depth = 10, min-samples-leaf = 2, max-features = 2 |
| Logistic Regression | LR | polynomial method = input method, outlier tolerance = 1.0E-8 |
| Artificial Neural Networks | ANN | multilayer perceptron, hidden layer = 2 |
| Support Vector Machines | SVM | Regularization parameter = 6, Kernel type = RBF, Gamma = 0.1 |
| Bayesian Network | BN | Structure type = TAN, Parameter learning = Bayesian adjustment |
| Decision Tree | DT | level of pruning = 0.85, min-samples-leaf = 2 |
| K-Nearest Neighbor | KNN | K = 4, Distance calculation = Euclidean metric |

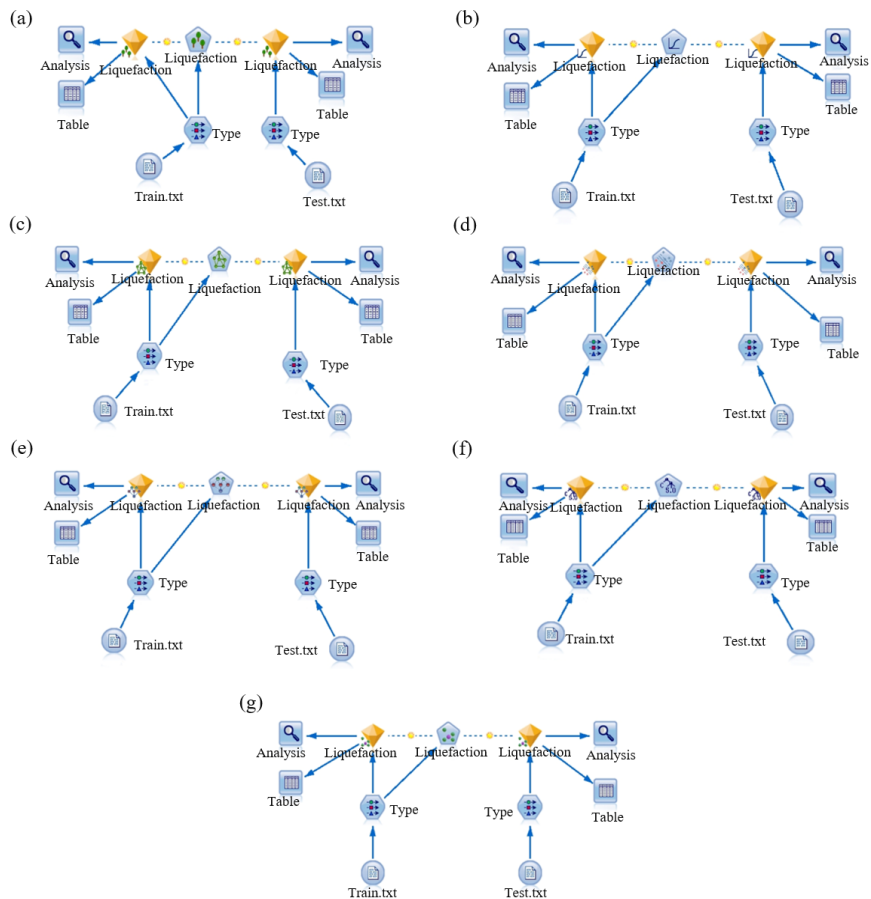

Figure 11. The models of seismic soil liquefaction: (a) RF model; (b) LR model; (c) ANN model; (d) SVM model; (e) BN model; (f) DT model; (g) KNN model.

## 4. The effect of sampling strategies on the performance of ML models

The learning performance and prediction performance of seven ML models under five sampling methods are shown in Figure 12 and Figure 13. From Figure 12, the mean values of the learning performance of the seven ML models under different sampling methods do not differ much, but there are differences in the learning performance of each model: the RF model always shows the optimal learning performance, with a maximum of $Acc$ = 98.1% and $F_1$ = 98.7%, and the DT model has the second highest learning performance. In contrast, the learning performance of the other five models is poor.

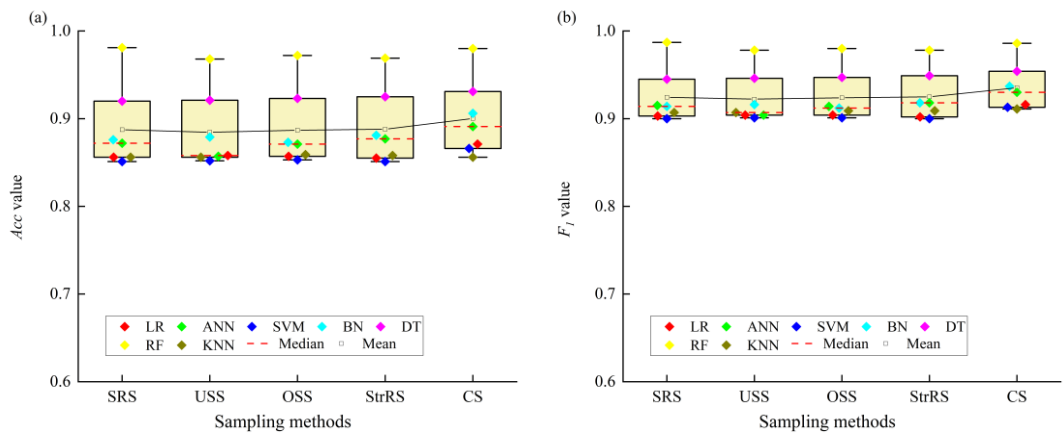

Figure 12. Comparison of the effect of sampling methods on model-learning performance: (a) $Acc$; (b) $F_1$.

From Figure 13, the mean values of the prediction performance of the seven ML models under different sampling methods differ significantly compared to the learning performance. Among them, under the OSS method, the prediction performance of all seven ML models is optimal, with the RF model performing the best and the KNN model the worst. Whereas, under the CS method, all seven ML models have the worst predictive performance. Therefore, it is determined that the optimal sampling method in this study is the OSS method.

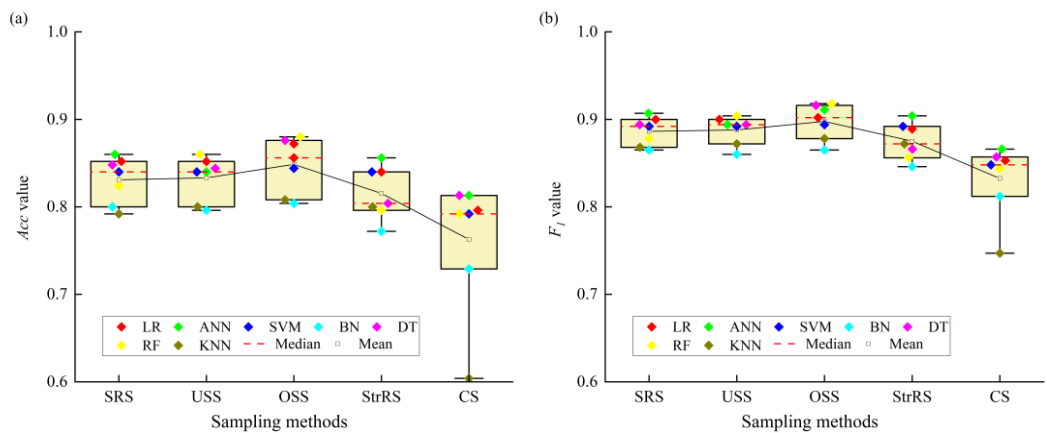

Figure 13. Comparison of the effect of sampling methods on model prediction performance: (a) $Acc$; (b) $F_1$.

To highlight the effect of sampling methods on the distribution of data sets, the best-performing sampling method (OSS) and the worst sampling method (CS) are selected in this study. Taking the best-performing RF model as an example, the distributions of each variable in the training and test sets under the two sampling methods are compared, as shown in Figure 14. From Figure 14, it can be seen that the distributions of each variable in the training and test sets under the OSS method are more similar. In order to further quantify the similarity of the distributions of the variables in the training and test sets under the two sampling methods, the KL divergence values are calculated using Eq. 17 and compared, as shown in Figure 15. It can be seen that the KL divergence values between variables under the OSS method are much smaller than those of the CS method. This indicates that the distribution between the training data and the test data is more similar using the OSS method compared to the CS method, suggesting better model performance using the OSS method. Therefore, the OSS method is chosen to study the effects of training set sample size, train-test split ratio, and class unbalance on the performance of seven ML models for predicting seismic liquefaction.

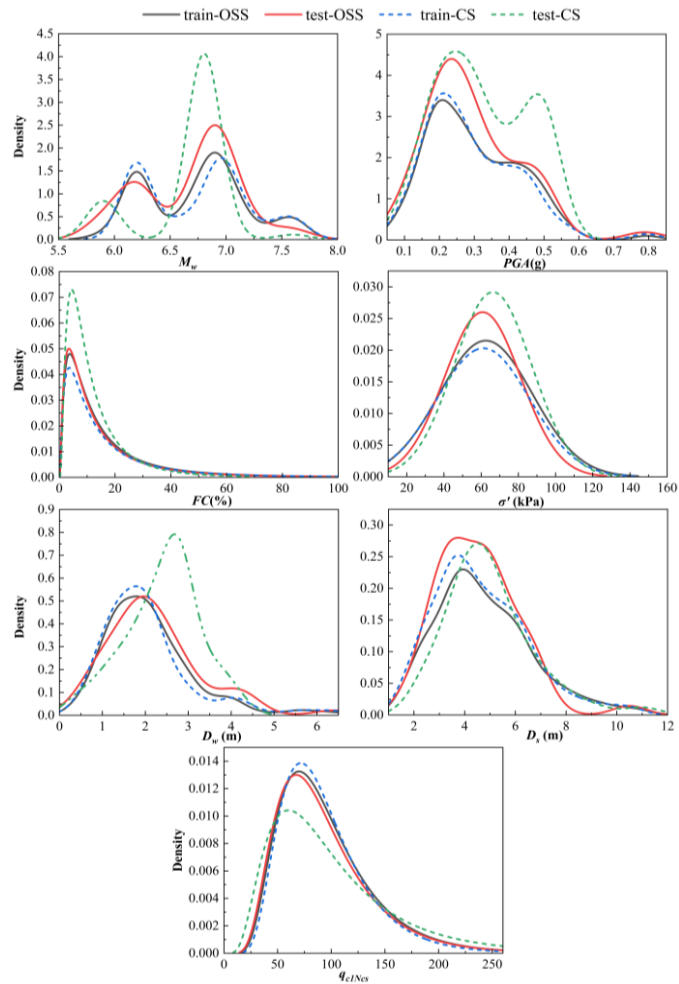

Figure 14. Distribution of variables in the training set and test set under OSS and CS methods.

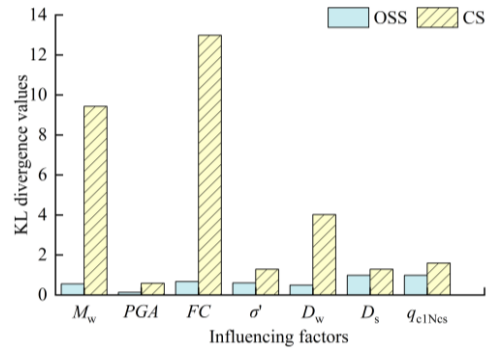

Figure 15. KL divergence values for each variable of the data under OSS and CS methods.

## 5. The effect of training set configurations on model performance under the OSS method

5.1 The effect of training set sample size on model performance

The learning performance and prediction performance of the models under the OSS method approach with different training set sample sizes (40, 80, 120, 160, and 200) are shown in Figure 16 and Figure 17. From Figure 16, it can be seen that the learning performance of the models shows an undulating upward trend as the training set sample size increases. When the training set sample size reaches 200, the learning performance of all seven ML models performs optimally, with the RF model performing the best and the SVM model performing the worst.

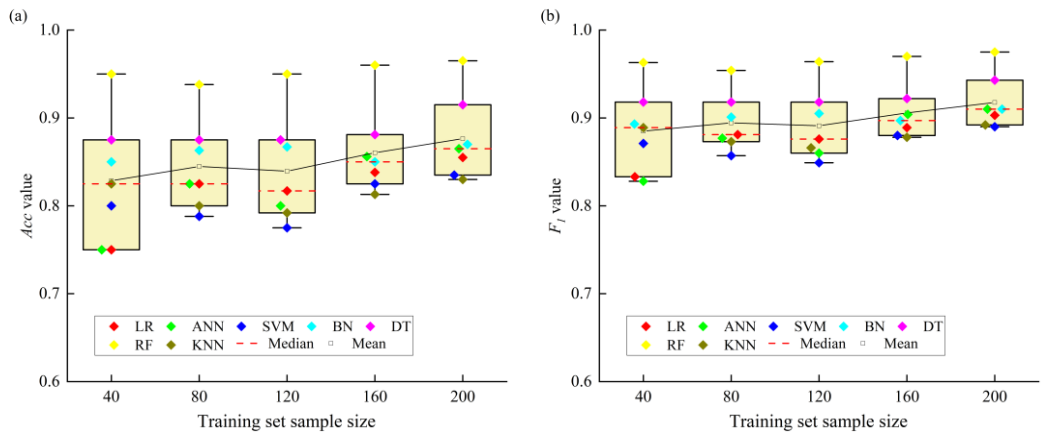

Figure 16. Comparison of the effect of training set sample sizes on model-learning performance: (a) $Acc$; (b) $F_1$.

From Figure 17, the prediction performance of each ML model under different training set sample sizes shows significant variability compared to the learning performance. When the training set sample size is 200, the prediction performance of all seven models is optimal, with the RF model performing the best and the DT model performing the worst. When the training set sample size is 40, the seven models have the worst prediction performance.

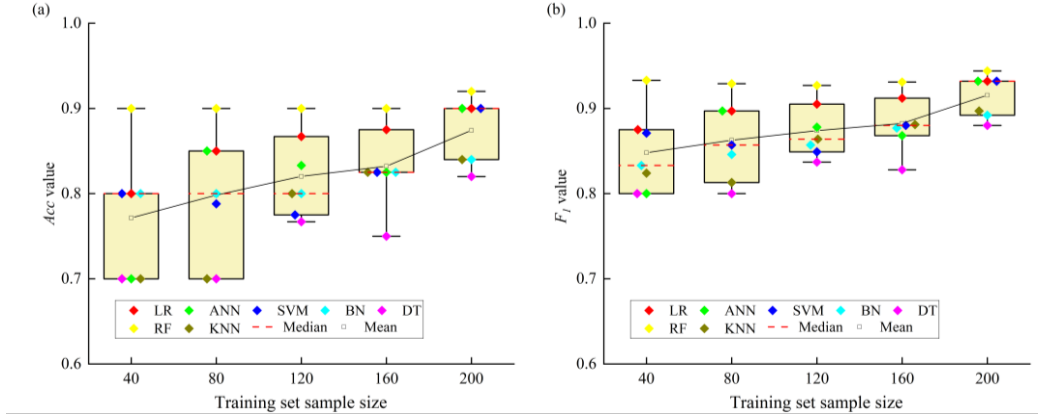

Figure 17. Comparison of the effect of training set sample sizes on model prediction performance: (a) *Acc*; (b) $F_1$.

Model performance improvement usually levels off after reaching a certain data size [17]. According to the results in Figure 18, the model performance improves with increasing sample size, and the optimal training set sample size cannot be determined. In order to further verify whether the model performance will stabilize with a larger sample size, the dataset is expanded to 350 samples. Eq. 23 is used to calculate the Average accuracy improvement (*AAI*) of the models to quantify the improvement in the performance of the seven machine learning models with different training set sample sizes.

$$AAI = \frac{\sum_{j=1}^{7}(y_{ij} - y_0)}{7} \tag{23}$$

where $y_{ij}$ ($i$=40, 80, 120, 160, 200, 240, or 280, $j$ is 7 ML models) is the *Acc* value of the $j$th model when the training set sample size is $i$, and $y_0$ is the *Acc* value of model $j$ when the training set sample size is 40.

The *AAI* of the model with different training set sample sizes is shown in Figure 18. With the increase of the training set sample size, the *AAI* of the model first shows an undulating upward trend. When the training set sample size increases to 200, the model's *AAI* reaches the maximum value. Subsequently, the *AAI* of the model leveled off when the training set sample size was increased from 200 to 280. Therefore, the optimal training set sample size of 200 was determined.

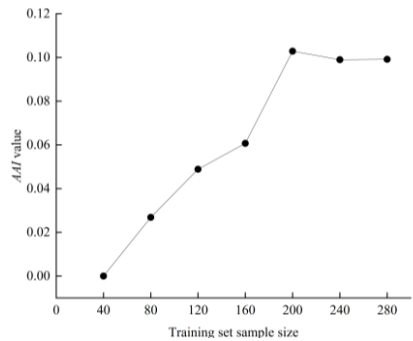

Figure 18. The effect of training set sample size on model *AAI*.

From Figure 18, when the training set sample size exceeds 200, the sample size does not have an impact on the model performance. In order to further compare the degree of influence on the prediction performance of the seven ML models when the training set sample size is 40, 80, 120, 160, and 200, the Influence degree ($I_d$) is calculated according to Eq. 24.

$$I_d = \frac{Acc_{max} - Acc_{min}}{Acc_{min}} \%$$ (24)

where $Acc_{max}$ is the maximum value of $Acc$ for a particular model with different training set sample sizes, and $Acc_{min}$ is the minimum value of $Acc$ for a particular model with different training set sample sizes.

The extent to which the training set sample size affects the performance of the seven ML models is shown in Table 7. From Table 7, it can be seen that there are differences in the $I_d$ of the training set sample size on model performance, with the training set sample size having the greatest influence on the RF model (31.4%), followed by the DT model (17.1%), and the smallest influence being on the KNN model (2.0%).

Table 7. Comparison of the effect of training set size on model performance.

| Model | Training sample size | $Acc_{max}$ | $Acc_{min}$ | $I_d$ |
|---|---|---|---|---|
| LR | 40, 80, 120, 160, and 200 | 90.0% | 80.0% | 12.5% |
| ANN | 40, 80, 120, 160, and 200 | 90.0% | 83.3% | 8.0% |
| SVM | 40, 80, 120, 160, and 200 | 90.0% | 77.5% | 16.1% |
| BN | 40, 80, 120, 160, and 200 | 84.0% | 80.0% | 5.0% |
| DT | 40, 80, 120, 160, and 200 | 82.0% | 70.0% | 17.1% |
| RF | 40, 80, 120, 160, and 200 | 92.0% | 70.0% | 31.4% |
| KNN | 40, 80, 120, 160, and 200 | 84.0% | 70.0% | 2.0% |

5.2 The effect of train-test split ratio on model performance

The learning performance and prediction performance of the models under five train-test split ratios (50:50, 60:40, 70:30, 80:20, and 90:10) under the OSS method are shown in Figure 19 and Figure 20. From Figure 19, it can be seen that different models have inconsistent sensitivity and adaptability to data. When the train-test split ratio is 50:50, the learning performance of the LR model and RF model is optimal. When the train-test split ratio is 60:40, the learning performance of the BN model is optimal. The learning performance of the ANN model, DT model, and KNN model is optimal when the train-test split ratio is 70:30. The learning performance of the SVM model is optimal when the train-test split ratio is 90:10.

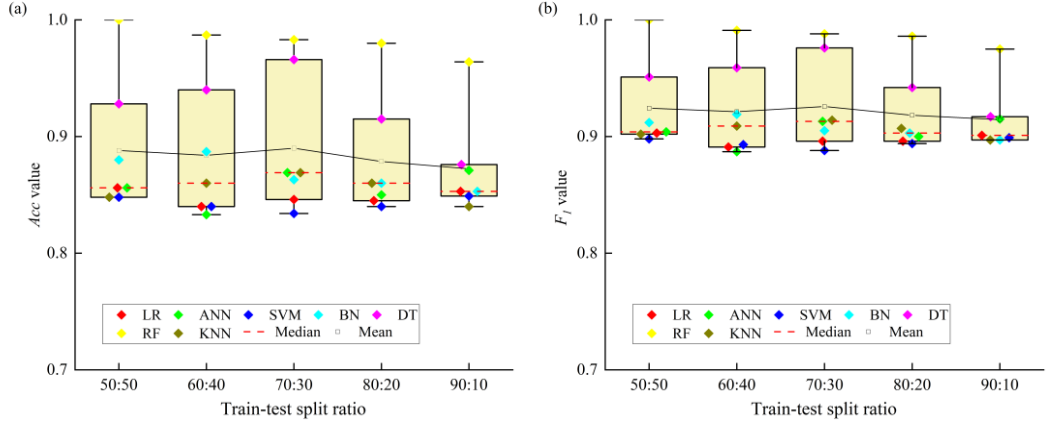

Figure 19. Comparison of the effect of train-test split ratios on model-learning performance: (a) $Acc$; (b) $F_1$.

From Figure 20, the prediction performance of each ML model with different train-test split ratios compared to the learning performance exhibits variability. For LR, ANN, and DT models, the predictive performance first decreases, then increases, and finally decreases as the train-test split ratio increases. For the SVM model, BN model, RF model, and KNN model, the prediction performance tends to increase and then decrease. When the train-test split ratio is 80:20, the prediction performance of all seven models is optimal, with the RF model performing the best and the KNN model performing the worst. When the train-test split ratio is 50:50, the seven models have the worst prediction performance. Therefore, the optimal train-test split ratio of 80:20 was determined for this study.

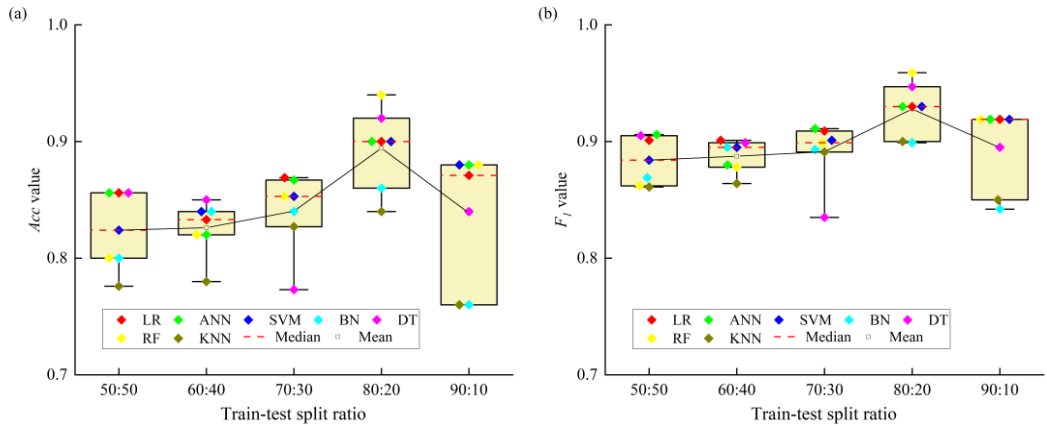

Figure 20. Comparison of the effect of train-test split ratios on model prediction performance: (a) $Acc$; (b) $F_1$.

In order to further compare the $I_d$ of the train-test split ratio on the prediction performance of the seven ML models, the $I_d$ was calculated according to Eq. 24, and the results were obtained as shown in Table 8. Table 8 reveals differential train-test split ratios impacts: strongest on DT (19.0%), followed by RF (17.5%), and weakest on BN (7.5%). Higher ratios improve DT through enhanced feature selection and noise resistance, while RF benefits from increased bootstrap diversity and reduced tree correlation. BN shows minimal sensitivity (5.0% training sample size impact), as Bayesian priors ensure small-data stability while diminishing returns occur with larger datasets.

Table 8. Comparison of the impact of train-test split ratio on ML model performance.

| Model | Train-test split ratio | $Acc_{max}$ | $Acc_{min}$ | $I_d$ |
| --- | --- | --- | --- | --- |
| LR | 50:50, 60:40, 70:30, 80:20, and 90:10 | 90.0% | 83.3% | 8.0% |
| ANN | 50:50, 60:40, 70:30, 80:20, and 90:10 | 90.0% | 82.0% | 9.8% |
| SVM | 50:50, 60:40, 70:30, 80:20, and 90:10 | 90.0% | 82.4% | 9.2% |
| BN | 50:50, 60:40, 70:30, 80:20, and 90:10 | 86.0% | 80.0% | 7.5% |
| DT | 50:50, 60:40, 70:30, 80:20, and 90:10 | 92.0% | 77.3% | 19.0% |
| RF | 50:50, 60:40, 70:30, 80:20, and 90:10 | 94.0% | 80.0% | 17.5% |
| KNN | 50:50, 60:40, 70:30, 80:20, and 90:10 | 84.0% | 76.0% | 10.5% |

5.3 The effect of class unbalance on model performance

The learning performance and prediction performance of the models under nine class distributions (20:80, 25:75, 33:67, 40:60, 50:50, 60:40, 67:33, 75:25, and 80:20) under the OSS method are shown in Figure 21 and Figure 22. From Figure 21(a), the worse the class imbalance is (from 50:50 to 20:80 or from 50:50 to 80:20), the larger the Acc value of the ANN model, LR model, and SVM model, while the change of Acc value of four other models (BN, DT, RF, and KNN) is not monotonically regular. From Figure 21(b), the F1 values of the seven ML models show an undulating upward trend as the class distribution increases, compared to the RF model, which does not change much in its learning performance.

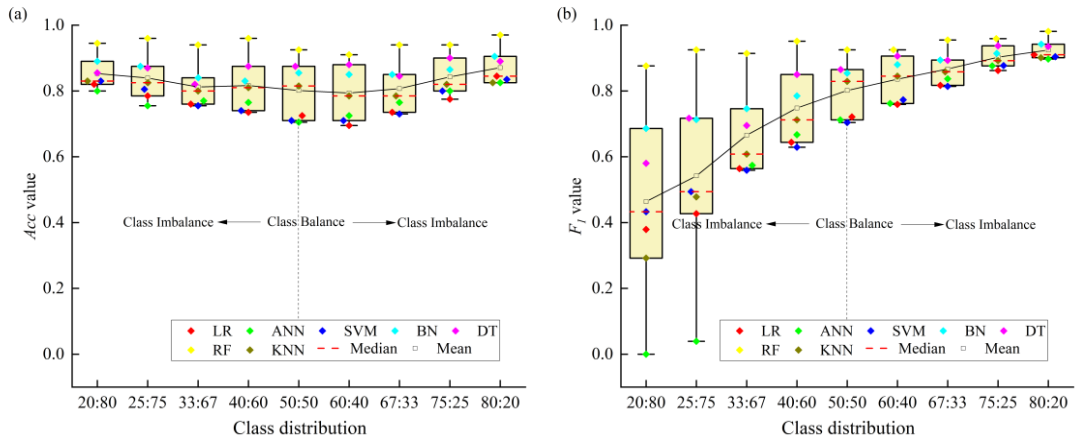

Figure 21. Comparison of the effect of class distributions on model-learning performance: (a)$Acc$; (b) $F_1$.

From Figure 22, SVM, BN, DT, RF, and KNN models have the best prediction performance when the class distribution is 60:40 (1.5:1). When the class distribution is 50:50 (1:1), the LR model and ANN model have the best prediction performance. It can be seen that the models show optimal prediction performance when the range of class distribution is 1 to 1.5.

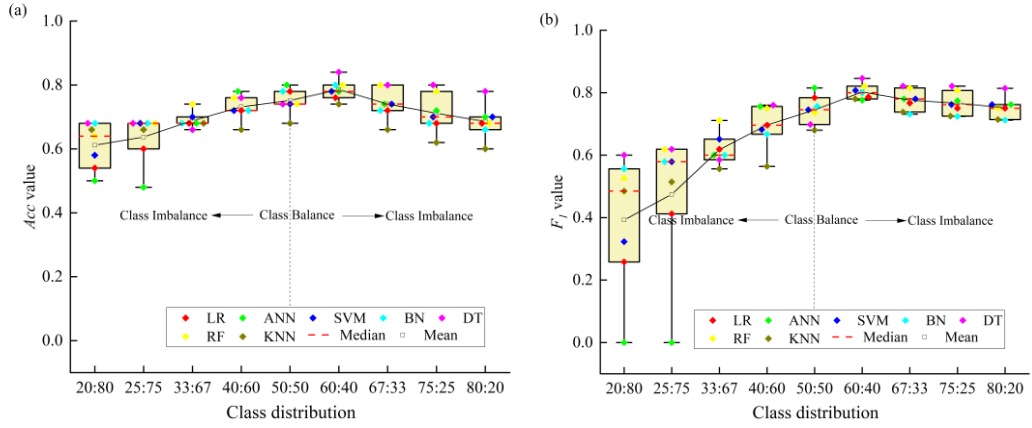

Figure 22. Comparison of the effect of class distributions on model prediction performance: (a) $Acc$; (b) $F_1$.

To further compare the $I_d$ of the class distributions on the seven models, the $I_d$ was calculated according to Eq. 6, and the results are shown in Table 9. From Table 9, there are differences in the $I_d$ of class distributions on model performance, with class distributions having the greatest $I_d$ on the ANN model (66.7%), followed by the LR model (44.4%), and the least influence on the KNN model (12.1%).

Table 9. Comparison of the impact of class distribution on ML model performance.

| Model | Class distribution | $Acc_{max}$ | $Acc_{min}$ | $I_d$ |
| --- | --- | --- | --- | --- |
| LR | 1:4, 1:3, 1:2, 1:1.5, 1:1, 1.5:1, 2:1, 3:1, and 4:1 | 78.0% | 54.0% | 44.4% |
| ANN | 1:4, 1:3, 1:2, 1:1.5, 1:1, 1.5:1, 2:1, 3:1, and 4:1 | 80.0% | 48.0% | 66.7% |
| SVM | 1:4, 1:3, 1:2, 1:1.5, 1:1, 1.5:1, 2:1, 3:1, and 4:1 | 78.0% | 58.0% | 34.5% |
| BN | 1:4, 1:3, 1:2, 1:1.5, 1:1, 1.5:1, 2:1, 3:1, and 4:1 | 80.0% | 68.0% | 17.6% |
| DT | 1:4, 1:3, 1:2, 1:1.5, 1:1, 1.5:1, 2:1, 3:1, and 4:1 | 84.0% | 66.0% | 27.3% |
| RF | 1:4, 1:3, 1:2, 1:1.5, 1:1, 1.5:1, 2:1, 3:1, and 4:1 | 80.0% | 64.0% | 25.0% |
| KNN | 1:4, 1:3, 1:2, 1:1.5, 1:1, 1.5:1, 2:1, 3:1, and 4:1 | 74.0% | 66.0% | 12.1% |

5.4 The effect of training set configurations coupling on model performance

Based on the previous study, the optimal training set sample size was determined to be 200, the optimal train-test split ratio was 80:20, and the optimal class distribution ranged from 1 to 1.5. In addition to the determined optimal training set configurations, two training set sample sizes, two train-test split ratios, and two class distributions with better performance were also selected. The effects of the coupled training set configurations (training set sample sizes: 120, 160, and 200; train-test split ratios: 60:40, 70:30, and 80:20; and class distributions: 50: 50, 60: 40, and 67: 33) on the model performance were investigated, and the results of the model's learning and prediction performance are shown in Figure 23 and Figure 24.

From Figure 23, the overall learning performance of the seven models has no significant upward or downward trend under different couplings of training set configurations. Among them, the RF model always maintains the optimal learning performance compared with other models, the BN model is the second best, and the learning performance of the LR model, ANN model, and SVM model is always poor.

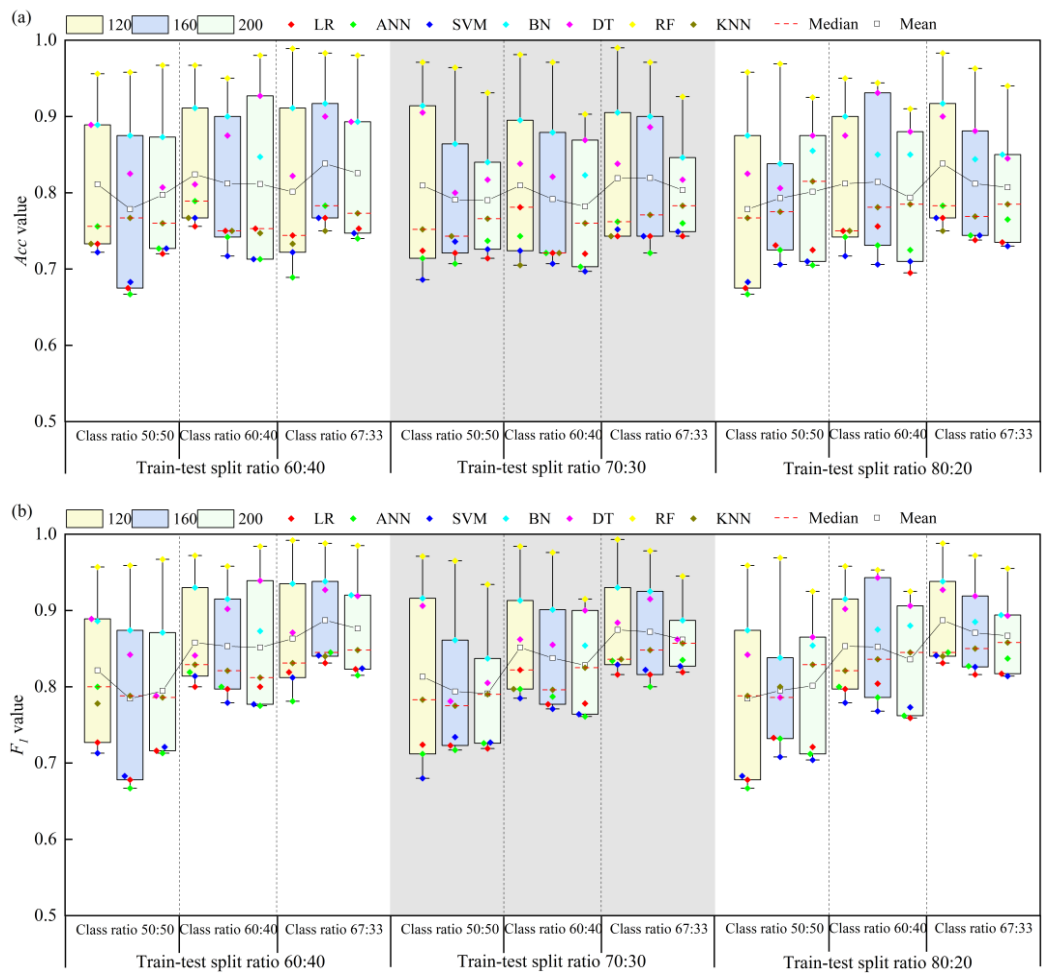

Figure 23. Comparison of the effect of training set configurations coupling on model-learning performance: (a) $Acc$; (b) $F_1$.

From Figure 24, the prediction performance of the seven ML models with different training set configurations coupling shows variability compared to the learning performance. When the training set sample size is 200, the train-test split ratio is 80:20, and the class distribution is 50:50 (1:1), the LR model has the best prediction performance. The predictive performance of the other six models (ANN, SVM, BN, DT, RF, and KNN) is optimal when the training set sample size is 200, the train-test split ratio is 80:20, and the class distribution is 60:40 (1.5:1).

Therefore, for the results of the effect of training set configurations coupling on model performance, the training set sample size of 200 and the train-test split ratio of 80:20 are recommended. For the LR model, the class distribution 50:50 is recommended. For the ANN, SVM, BN, DT, RF, and KNN models, the class distribution of 60:40 is recommended.

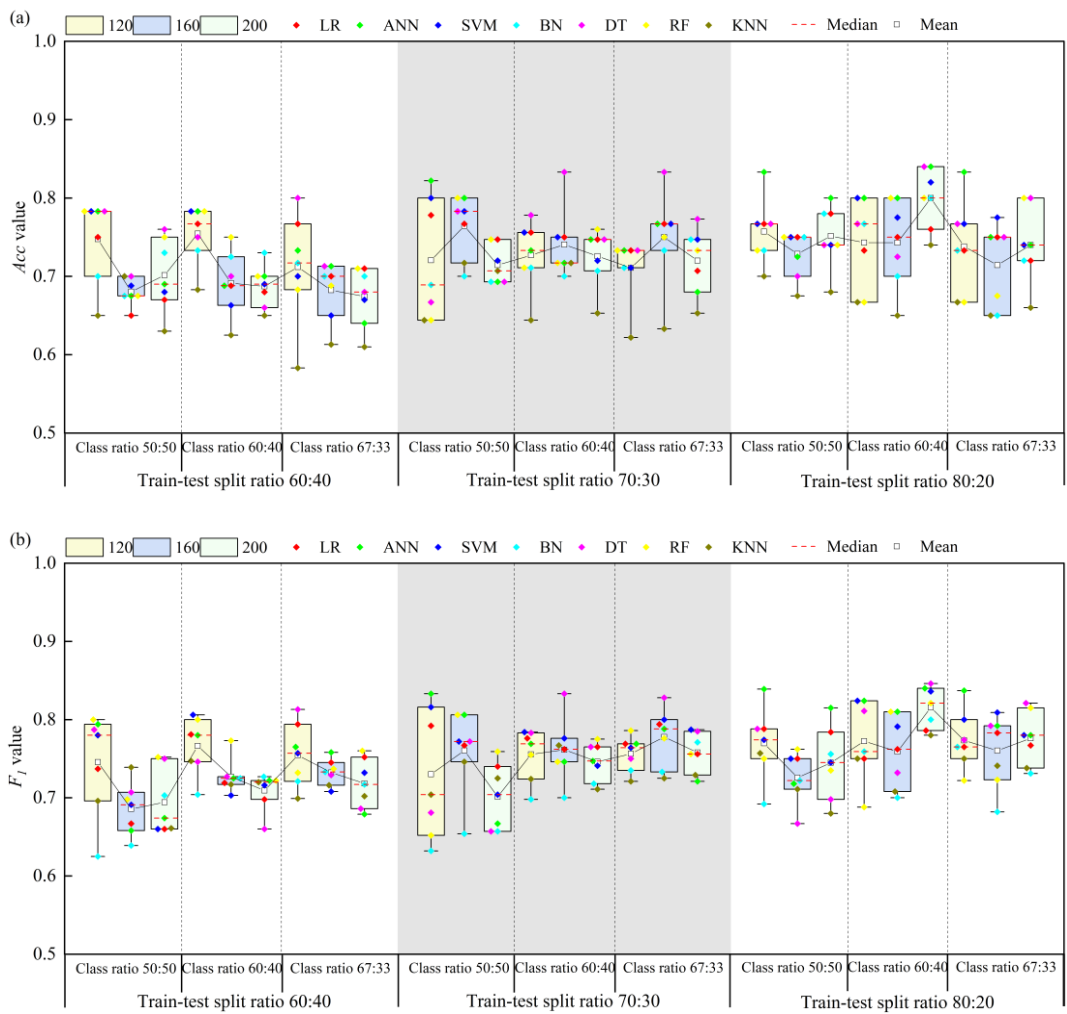

Figure 24. Comparison of the effect of training set configurations coupling on model prediction performance: (a) $Acc$; (b) $F_1$.

To further compare $I_d$ of the training set configurations coupling on the prediction performance of the seven ML models, the results of $I_d$ are obtained as shown in Table 9, calculated according to Eq. 24. From Table 9, the training set sample size has a higher $I_d$ on the performance of the BN model, DT model,

RF model, and KNN model, which are all more than 13%, in contrast to the training set sample size, which has a smaller $I_d$ on the performance of LR model, ANN model, and SVM model, which is not more than 6%. The train-test split ratio has a high impact on all seven models, which is more than 12%, and the impact is more than 20% for the ANN and DT models. Compared with the training set sample size and the train-test split ratio, the class distribution has a lower impact on the seven models. For LR, ANN, SVM, and DT models, the train-test split ratio has the greatest influence. For the BN model, RF model, and KNN model, the training set sample size has the greatest effect on them.

In terms of the average $I_d$ of the three training set configurations on model performance, the train-test split ratio has the greatest $I_d$, the class distribution is the next largest, and the training set sample size has the least $I_d$. The smallest impact of the training set sample size is because the minimum sample size chosen in this section is 120, which is a certain size and therefore has less impact on the model. The $I_d$ of the class distribution is also small because the selected class distribution is controlled between 1 and 2, so the $I_d$ is also small. The train-test split ratio directly affects the test set sample size, which will have a great impact on the prediction performance. The train-test split ratio directly determines whether the model can guarantee good learning performance and good prediction performance at the same time, and is the baseline guarantee of performance. The class distribution, on the other hand, indirectly interferes with the model decision through distributional bias but can be partially adjusted by the algorithm. The marginal benefit of the training set sample size and the difference in model data requirements minimize the impact in non-extreme cases.

Table 10. Comparison of the $I_d$ of training set configurations on model performance.

| Model | $I_d$ of training set sample size | $I_d$ of train-test split ratio | $I_d$ of class distribution |
| --- | --- | --- | --- |
| LR | 13.90% | 57.10% | 28.90% |
| ANN | 13.00% | 52.00% | 35.10% |
| SVM | 16.40% | 53.10% | 30.50% |
| BN | 37.10% | 34.20% | 28.80% |
| DT | 28.00% | 48.10% | 23.80% |
| RF | 47.00% | 33.80% | 19.10% |
| KNN | 34.80% | 30.50% | 34.80% |
| Mean | 27.20% | 44.10% | 28.70% |

## 6. Discussion

To provide a systematic and data-centric foundation for reliable machine-learning-based seismic liquefaction prediction, this study establishes a comprehensive multi-model benchmark to evaluate the influence of sampling strategies (SRS, USS, OSS, StrRS, and CS), training set sample sizes (40, 80, 120, 160, and 200), train-test split ratios (50:50, 60:40, 70:30, 80:20, and 90:10), class distributions (20:80, 25:75, 33:67, 40:60, 50:50, 60:40, 67:33, 75:25, and 80:20), and training set configurations coupling on the performance of seven ML models (LR, ANN, SVM, BN, DT, RF, and KNN) for predicting seismic liquefaction. The benchmark results demonstrate that the OSS method is the optimal sampling method. Then, based on the OSS method, the optimal training set sample size of 200, the optimal train-test split ratio of 80:20, and the optimal class distribution range of 1~1.5 are determined.

The datasets obtained from different sampling methods differ in distribution [30], thus affecting the model performance to varying degrees. Demir and Şahin [11] showed that the StrRS method outperforms the SRS method for SVM, RF, and XGBoost models. Therefore, Zhou et al. [10] and Demir and Şahin [16], [31] directly adopted StrRS without considering the effect of different sampling methods on model performance. Hu et al. [51] previously demonstrated that within RF models, the OSS method yields better predictive performance than other probabilistic sampling methods. This study further confirms that the superiority of the OSS method also applies to other machine learning algorithms. Therefore, based on these studies, the OSS, USS, and CS methods are added to investigate the effect of different sampling methods on model performance, and the optimal sampling method is the OSS method. The results of the OSS method are better than those of the StrRS method, because of the double sorting mechanism in the OSS method (the main category is sorted according to whether liquefaction, and then the detailed sorting is done according to the weight of seismic influencing factors) together with the fixed-interval sampling make the samples more balanced in the feature space, which can effectively cover the different combinations of features and avoid sampling bias due to the clustering of extreme values or the class imbalance, and maximize the similarity between the distribution of the training set and the test set, as shown in Figure 14. In contrast, the StrRS method is only based on a single variable and fails to capture the complex interactions between multiple variables, resulting in the model easily learning "fragmented" features and reduced generalization ability.

To further verify the accuracy of the double ordering mechanism in the OSS method, the variable ordering methods in this study were compared with other scholars, as shown in Table 11. Figure 25 compares the average performance of the seven models under the nine variable ordering methods in Table 11. The performance of the model under the ordering approach proposed in this study is better than the other ordering approaches. Different variable ordering methods reflect distinct emphases on the liquefaction mechanism, directly influencing the priority of model input parameters and leading to performance disparities: Xue and Xiao [32] weakens the influence of *PGA*, which leads to the under-distribution of high *PGA* samples in the training set, and the model's poor ability to generalize to the strong earthquake scenario. Ghani et al. [33] takes *FC* as the last variable, and its threshold effect may be ignored. And $\sigma'$ is ranked later, neglecting its strong coupling with $D_s$ and $D_w$. This study prioritizes $q_{c1Ncs}$ and *PGA* in the variable ordering to comprehensively account for soil liquefaction resistance and ground shaking intensity [34], to ensure that the training set and the test set have the same proportion of samples in the high-liquefaction-risk zone (low $q_{c1Ncs}$ + high *PGA*) and the low-liquefaction-risk zone (high $q_{c1Ncs}$ + low *PGA*), and to avoid the model being over- or under fitted due to the bias of the data delineation. The subsequent ordering of $\sigma'$, *FC*, ensures that samples with different overburden stresses and fine-grained contents are equally divided under the same liquefaction resistance and ground shaking intensity, which improves the generalization ability of the model to complex working conditions and significantly improves the physical reasonableness of liquefaction discrimination. The accuracy of ordered systematic sampling can be seen.

Table 11. Ordering of the importance of influencing factors in the literature.

| Literature | Model or algorithm | The order of importance of influencing factors |
| --- | --- | --- |
| This study | LR, ANN, SVM, BN, DT, RF, and KNN | $q_{c1Ncs}$→*PGA*→$\sigma'$→*FC*→$D_s$→$M_w$→$D_w$ |
| Xue and Xiao [32] | SVM | *FC*→$D_s$→$\sigma'$→$(N_1)_{60,cs}$→*PGA*→$D_w$→$M_w$ |
| Cai et al. [35] | SVM, ANN | $(N_1)_{60,cs}$→*FC*→$D_w$→*PGA*→$\sigma'$→$D_s$→$M_w$ |
| Zhou et al. [36] | SVM | $M_w$→*PGA*→*FC*→$\sigma'$→$D_w$→$D_s$→$q_{c1Ncs}$ |
| Chern et al. [37] | ANN | *FC*→$(N_1)_{60,cs}$→$D_w$→$\sigma'$→$D_s$→*PGA*→$M_w$ |
| Mustafa et al. [38] | RF, DT, KNN, and SVM | *FC*→$(N_1)_{60,cs}$→$D_s$→$\sigma'$→$D_w$→*PGA*→$M_w$ |
| Ghani et al. [33] | DT, RF | $(N_1)_{60,cs}$→*PGA*→$D_s$→$M_w$→$\sigma'$→$D_w$→*FC* |
| Kumar et al. [39] | ANN | $M_w$→*PGA*→*FC*→$\sigma'$→$D_w$→$(N_1)_{60,cs}$→$D_s$ |
| Jas and Dodagoudar [40] | XGBoost | $q_{c1Ncs}$→$D_w$→*PGA*→*FC*→$D_s$→$\sigma'$→$M_w$ |

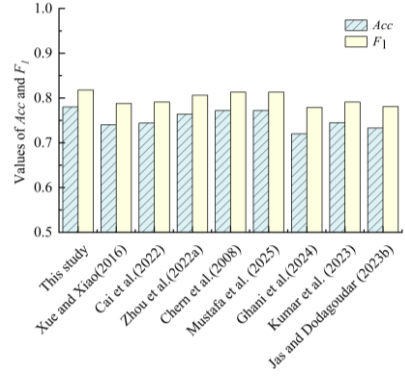

Figure 25. Comparison of model performance of different ordering methods in the literature.

Existing studies [17], [41]-[43] have verified that there is not a simple linear relationship between the improvement of model performance and the increase of the sample size of the training set: the model performance will improve with the increase of the sample size of the training set, and the improvement of the model performance will gradually level off when the size of the sample size is increased to a certain degree. This law is also derived from this study. However, most of the existing studies are limited to a single model or only stay at the stage of qualitatively describing the relationship between sample size and model performance, without further quantitatively determining the optimal sample size. In this study, we compare the performance of seven commonly used machine learning models for predicting seismic liquefaction and determine the optimal sample size of the training set to be 200, to further verify the accuracy of the results, we calculate the $H(X, Y)$ of the input and output variables in the training set samples according to Eq. 17, as well as the incremental increase of the joint entropy $\Delta H(X, Y)$ (according to Eq. 25).

$$\Delta H(X,Y) = H_{i+40}(X,Y) - H_i(X,Y) \tag{25}$$

where $H_i(X, Y)$ denotes the value of sample joint entropy when the sample size of the training set is i (i = 40, 80, 120, 160, 200, and 240). The calculation results are shown in Figure 26. From Figure 26 (a), it can be seen that the joint information entropy $H(X, Y)$ tends to increase as the sample size increases. However, from Figure 26 (b), it can be seen that the increment of joint entropy $\Delta H(X, Y)$ shows an overall decreasing trend. When the training sample size reaches 200, the information entropy increase slows down significantly ($\Delta H(X, Y)$ decreases to 0.24 from 0.43 in the previous stage), and thereafter, even though the sample size of the training set is increasing, $H(X, Y)$ stays around 0.23, indicating that the added information that can be provided by new samples gradually tends to stabilize. At the same time, the prediction performance of the seven machine learning models peaks at 200 samples in the training set and tends to stabilize (Figure 18), and the joint entropy and model performance curves jointly show an inflection point, indicating that the new samples mainly bring redundant rather than effective information. In summary, the training set sample size of 200 is optimal for both information and performance saturation.

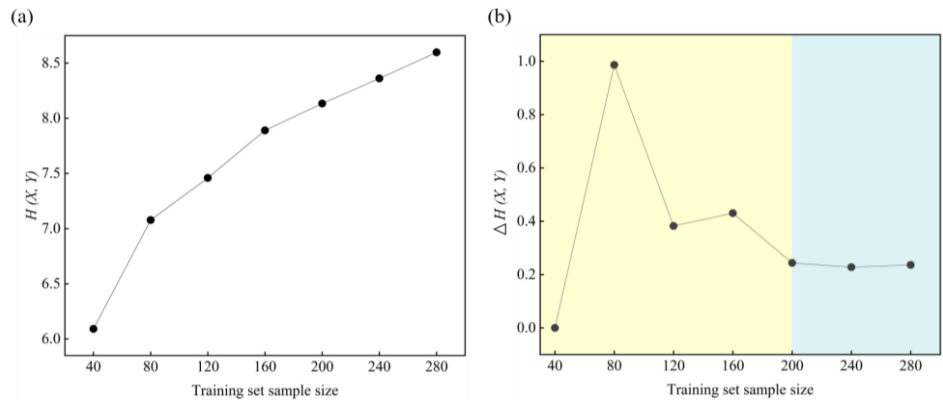

Figure 26. Comparison of joint entropy of training set sample size: (a) $H(X, Y)$; (b) $\Delta H(X, Y)$.

Currently, most scholars adopt the 80:20 train-test split ratio [44]-[48], and this study also verifies that this ratio can obtain a better model performance in predicting seismic liquefaction. This result is in line with Pareto's principle (i.e., the "eight-two rule"): 80% of the samples are used as the training set, which provides sufficient learning resources for the model, and 20% of the samples are retained as the test set, which ensures the objectivity and generalization of the evaluation. In essence, it is the result of multiple mechanisms working together: a quantitative balance of statistical variance, information entropy saturation, and model complexity is achieved at a ratio of 80:20, thus becoming co-optimal in terms of performance, robustness, and information utilization. Unlike the existing literature, which focuses on a single model, this paper systematically compares the performance of seven commonly used machine learning models for predicting earthquake liquefaction and determines the optimal train-test split ratio of 80:20, which is more generalizable.

To address the problem that there is a class imbalance in the dataset, Cetin et al. [49] used two approaches: the first approach was to suggest a range of class distribution from 1.5 to 3 by using expert interviews. The second approach was to consider the class proportions as unknown parameters in the model, which resulted in the optimal class distribution of 1.5. Hu et al. [28] pointed out that the optimal class distributions varied from one model to the next (BN: 1~1.5; ANN: 0.67~1; LR: 0.5; SVM: 0.5~1). In this study, we added LR, DT, and RF models to these scholars' studies investigated the effect of class imbalance on the performance of seven commonly used machine learning models in predicting seismic liquefaction, and came up with the optimal class distribution range of 1~1.5, which is somewhat different from the results of Hu et al. [28]. This is because Hu's study ignored the effect of different sampling methods on model performance, and the data sets obtained from different sampling methods differ in distribution. In this study, we fully consider the effects of different sampling methods on model performance and conduct a study of class imbalance based on optimal ordered system sampling, and the resulting range of class distribution is more universal. When the class distribution range is 1~1.5, the distribution of liquefied and non-liquefied samples in the dataset is more balanced, which improves the learning ability and generalization performance of the model. As the class distribution shifts towards the extremes (e.g., 20:80 or 80:20), the number of liquefied samples significantly decreases or increases, and the class distribution becomes unbalanced.

Besides, Hu et al. [50], Nguyen et al. [15], and Cetin et al. [49] suggested that sample size, the train-test split ratio, and class unbalance affect the model performance, but no one has investigated the effect of the coupling of these training set configurations on the model performance. In this study, we first explore the impact of training set configurations coupling on the performance of machine learning models and quantify the impact of different training set configurations on the performance of different ML models, which fills the gap in the current research.

Although this study provides a meaningful reference for choosing appropriate sampling methods and training set conditions, the following shortcomings still exist: (1) The sampling methods chosen in this paper are only probabilistic sampling methods, and the effects of self-help, oversampling, and undersampling on the performance of liquefaction discrimination models can be investigated later. (2) In this study, we have chosen seven common machine learning models for predicting seismic liquefaction. These seven machine learning algorithms are all supervised algorithms, and unsupervised learning, semi-supervised learning, and reinforcement learning can be studied in the future.

## 7. Conclusion

This study presents a comprehensive multi-model benchmark to systematically evaluate the influence of sampling strategies (SRS, USS, OSS, StrRS, and CS), training set sample sizes (40, 80, 120, 160, and 200), train-test split ratios (50:50, 60:40, 70:30, 80:20, and 90:10), and class distributions (20:80, 25:75, 33:67, 40:60, 50:50, 60:40, 67:33, 75:25, and 80:20) on the performance of seven ML models (LR, ANN, SVM, BN, DT, RF, and KNN) for predicting seismic liquefaction, and the following conclusions

can be drawn:

(1) The predictive performance of the seven ML models is optimal under the OSS method, and the predictive performance of the seven models is the worst under the CS method. The distributions between the training data and the test data using the OSS method are more similar.

(2) The predictive performance of the seven ML models is optimal when the training set sample size is 200, and the predictive performance of the seven models is worse when the training set sample size is 40. The training set sample size has the greatest degree of influence on the RF model and the least degree of influence on the KNN model.

(3) When the train-test split ratio is 80:20, the prediction performance of the seven ML models is optimal; when the train-test split ratio is 50:50, the prediction performance of the seven models is the worst. The train-test split ratio has the greatest influence on the DT model and the least influence on the BN model.

(4) The seven ML models exhibit optimal prediction performance when the class distribution in the training set ranges from 1 to 1.5. The class distribution has the greatest degree of influence on the ANN model and the least on the KNN model.

(5) When the training set sample size is 200, the train-test split ratio is 80:20, and the class distribution ranges from 1 to 1.5, all seven ML models show optimal prediction performance. The train-test split ratio has the greatest effect on the LR, ANN, SVM, and DT models; the training set sample size has the greatest effect on the BN, RF, and KNN models. In particular, from the average of the prediction performance results of the seven models, the train-test split ratio has the greatest influence on models, followed by the class distribution, and the training set sample size has the least influence.

**CRediT authorship contribution statement**

J.L. Hu: Writing – review & editing, Conceptualization, Resources, Supervision, Validation, Funding acquisition; F.L. He: Writing – original draft, Methodology, Code; L.M. Huang: Writing – original draft, Methodology, Code; Qianfeng Wang: Writing – review & editing.

**Declarations**

**Competing Interests** The authors declare that they have no known competing financial interests or personal relationships that could have influenced the work reported in this paper.

**Ethical and Informed Consent for Data Used** This study does not involve human participants or animals. A statement of informed consent is not applicable because the manuscript does not contain any personally identifiable information or personal data.

**Data availability**

Data will be made available on request.

**Acknowledgments**

This work was supported by the Natural Science Foundation of Hubei Province of China (No. 2025AFB567), the Foundation of Key Laboratory of Soft Soils and Geoenvironmental Engineering (Zhejiang University), Ministry of Education (No. 2025P03), and the China Three Gorges University of Research and Innovation Foundation Funded Project (No. 2024SSCX006).

**References:**

1. Li, Z.Y., Zhang, S., Yuan, X.M., et al. (2024). Characteristics of disasters caused by the Jishishan Ms 6.2 earthquake in Gansu Province in 2023. Journal of Institute of Disaster Prevention, 26(2):43-52. In Chinese. https://doi.org/10.3969/j.issn.1673-8047.2024.02.005.

2. Şehmusoğlu, E.H., Kurnaz, T.F., Erden, C. (2025). Estimation of soil liquefaction using artificial intelligence techniques: an extended comparison between machine and deep learning approaches. Environ. Earth Sci, 84(5), 1-22. https://doi.org/10.1007/s12665-025-12116-4.

3. Shen, M., Chen, Q., Zhang, J., Gong, W., Hsein Juang, C. (2016). Predicting liquefaction probability based on shear wave velocity: an update. Bull. Eng. Geol. Environ, 75, 1199-1214. https://doi.org/10.1007/s10064-016-0880-8.

4. Goh, A.T. (2002). Probabilistic neural network for evaluating seismic liquefaction potential. Can. Geotech.J, 39(1), 219-232. https://doi.org/10.1139/t01-073.

5. Singh, S.V., Ghani, S. (2024). A smarter approach to liquefaction risk: harnessing dynamic cone penetration test data and machine learning for safer infrastructure. Front Built Environ, 10, 1495472. https://doi.org/10.3389/fbuil.2024.1495472.

6. Hu, J.L., Liu, H.B. (2019). Bayesian network models for probabilistic evaluation of earthquake-induced liquefaction based on CPT and Vs databases. Eng. Geol, 254, 76-88. https://doi.org/10.1016/j.enggeo.2019.04.003.

7. Kohestani, V.R., Hassanlourad, M., Ardakani, A. (2015). Evaluation of liquefaction potential based on CPT data using random forest. Nat. Hazard, 79(2), 1079-1089. https://doi.org/10.1007/s11069-015-1893-5.

8. Hu, J.L., Liu, H.B. (2020). Comparison of data-driven methods for evaluating earthquake-i

nduced liquefaction potential. In book: Information Technology in Geo-Engineering, Proceedings of the 3rd International Conference (ICITG), Guimar˜aes, Portugal (pp. 353-364). https://doi.org/10.1007/978-3-030-32029-4_30.

9.  Nejad, A.S., Güler, E., Özturan, M. (2018). Evaluation of liquefaction potential using random forest method and shear wave velocity results. In: Proceedings of the 2018 International Conference on Applied Mathematics and Computational Science (pp. 23-233). https://doi.org/10.1109/ICAMCS.NET46018.2018.00012.

10. Zhou, J., Huang, S., Zhou, T., Armaghani, D.J., Qiu, Y. (2022b). Employing a genetic algorithm and grey wolf optimizer for optimizing RF models to evaluate soil liquefaction potential. Artif Intell Rev, 55(7), 5673-5705. https://doi.org/10.1007/s10462-022-10140-5.

11. Demir, S., Şahin, E.K. (2022a). Liquefaction prediction with robust machine learning algorithms (SVM, RF, and XGBoost) supported by genetic algorithm-based feature selection and parameter optimization from the perspective of data processing. Environ. Earth Sci, 81, 459. https://doi.org/10.1007/s12665-022-10578-4.

12. Alobaidi, M.H., Meguid, M.A., Chebana, F. (2019). Predicting seismic-induced liquefaction through ensemble learning frameworks. Sci. Rep, 9, 11786. https://doi.org/10.1038/s41598-019-48044-0.

13. Hoang, N.D., Bui, D.T. (2018). Predicting earthquake-induced soil liquefaction based on a hybridization of kernel Fisher discriminant analysis and a least squares support vector machine: a multi-dataset study. Bull. Eng. Geol. Environ, 77, 191-204. https://doi.org/10.1007/s10064-016-0924-0.

14. Jas K., Dodagoudar G.R. (2023a). Liquefaction potential assessment of soils using machine learning techniques: a state-of-the-art review from 1994-2021. Int. J. Geomech, 23(7): 03123002. https://doi.org/10.1061/IJGNAI.GMENG-7788.

15. Nguyen Q.H., Ly, H. B., Ho, L. S., et al. (2021). Influence of data splitting on performance of machine learning models in prediction of shear strength of soil. Mathematical Problems in Engineering, 2021(1), 4832864. https://doi.org/2021(1): 4832864. 10.1155/2021/4832864.

16. Demir, S., Sahin, E.K. (2022b). Comparison of tree-based machine learning algorithms for predicting liquefaction potential using canonical correlation forest, rotation forest, and random forest based on CPT data. Soil Dyn. Earthquake Eng, 154, 107130.
36

https://doi.org/10.1016/j.soildyn.2021.107130.

17. Hu J.L., Zou W., Wang J., et al. (2021). Minimum training sample size requirements for achieving high prediction accuracy with the BN model: A case study regarding seismic liquefaction. Expert Systems with Applications, 185, 115702. https://doi.org/10.1016/j.eswa.2021.115702.

18. Ahmad, M., Tang, X.W., Qiu, J.N., Ahmad, F. (2019). Evaluating Seismic Soil Liquefaction Potential Using Bayesian Belief Network and C4.5 Decision Tree Approaches. Appl. Sci, 9, 4226. https://doi.org/10.3390/app9204226.

19. Ahmad, M., Tang, X.W., Qiu, J.N., Ahmad, F., Gu, W.J. (2021). Application of machine learning algorithms for the evaluation of seismic soil liquefaction potential. Front. Struct. Civ. Eng, 15, 490–505. https://doi.org/10.1007/s11709-020-0669-5.

20. Jas, K., Mangalathu, S., Dodagoudar, G.R. (2024). Evaluation and analysis of liquefaction potential of gravelly soils using explainable probabilistic machine learning model. Comput. Geotech, 167, 106051. https://doi.org/10.1016/j.compgeo.2023.106051.

21. Bherde, V., Gorantala, N., & Balunaini, U. (2024). Liquefaction susceptibility prediction using ML-based voting ensemble classifier. Natural Hazards, 1-26. https://doi.org/10.1007/s11069-024-06960-z.

22. Chithuloori, P., & Kim, J. M. (2025). Soft voting ensemble classifier for liquefaction prediction based on SPT data. Artificial Intelligence Review, 58(8), 228. https://doi.org/10.1007/s10462-025-11230-w.

23. Fadliansyah, F., Faris, F., Wilopo, W. (2024). Implementation of machine learning classification models considering the optimum data ratio in predicting soil liquefaction susceptibility. IOP Conference Series: Earth and Environmental Scienc, 1416(1): 012012. https://doi.org/10.1088/1755-1315/1416/1/012012.

24. Ghani, S., Thapa, I., Kumari, S., et al. (2025). Revealing the nature of soil liquefaction using machine learning. Earth Science Informatics, 18(2), 198. https://doi.org/10.1007/s12145-024-01688-7.

25. Zhang, Y., Xie, Y., Zhang, Y., Qiu, J., Wu, S. (2021). The adoption of deep neural network (DNN) to the prediction of soil liquefaction based on shear wave velocity. Bulletin of Engineering Geology and the Environment, 80, 5053-5060. https://doi.org/10.1007/s10064-0

21-02250-1.

26. Oommen, T., Baise, L.G., Vogel, R.M. (2011). Sampling bias and class imbalance in maximum-likelihood logistic regression. Math. Geosci, 43, 99-120. https://doi.org/10.1007/s11004-010-9311-8.

27. Jain, A. (2012). Sampling bias in evaluating the probability of seismically induced soil liquefaction with SPT & CPT case histories. Michigan Technological University. https://doi.org/10.37099/mtu.dc.etds/459.

28. Hu, J.L., Tang, X.W., Qiu, J.N. (2017). Analysis of the influences of sampling bias and class imbalance on performances of probabilistic liquefaction models. Can. Geotech. J, 17(6), 04016134. https://doi.org/10.1061/(ASCE)GM.1943-5622.0000808.

29. Jena, R., Pradhan, B., Almazroui, M., Assiri, M., Park, H.J. (2023). Earthquake-induced liquefaction hazard mapping at national-scale in Australia using deep learning techniques. Geosci. Front, 14(1), 101460. https://doi.org/10.1016/j.gsf.2022.101460.

30. Zhou, L. (2013). Performance of corporate bankruptcy prediction models on imbalanced dataset: The effect of sampling methods. KBS, 41, 16-25. https://doi.org/10.1016/j.knosys.2012.12.007.

31. Demir, S., Sahin, E.K. (2023). An investigation of feature selection methods for soil liquefaction prediction based on tree-based ensemble algorithms using AdaBoost, gradient boosting, and XGBoost. Neural Comput. Appl, 35(4), 3173-3190. https://doi.org/10.1007/s00521-022-07856-4.

32. Xue X., Xiao M. (2016). Application of genetic algorithm-based support vector machines for prediction of soil liquefaction. Environ. Earth Sci, 75, 1-11. https://doi.org/10.1007/s12665-016-5673-7.

33. Ghani S., Sapkota S.C., Singh R.K., et al. (2024). Modelling and validation of liquefaction potential index of fine-grained soils using ensemble learning paradigms. Soil Dyn. Earthquake Eng, 177, 108399. https://doi.org/10.1016/j.soildyn.2023.108399.

34. Robertson P.K. (2015). Comparing CPT and Vs liquefaction triggering methods. J. Geotech. Geoenviron. Eng, 141(9), 04015037. https://doi.org/10.1061/(ASCE)GT.1943-5606.0001338.

35. Cai, M., Hocine, O., Mohammed, A.S., Chen, X., Amar, M.N., Hasanipanah, M. (2022). Integrating the LSSVM and RBFNN models with three optimization algorithms to predict the soil liquefaction

potential. Eng. Comput, 38, 3611–3623. https://doi.org/10.1007/s00366-021-01392-w.

36. Zhou J., Huang S., Wang M., et al. (2022a). Performance evaluation of hybrid GA-SVM and GWO-SVM models to predict earthquake-induced liquefaction potential of soil: a multi-dataset investigation. Eng Comput-germany, 1-19. https://doi.org/10.1007/s00366-021-01418-3.

37. Chern, S.G., Lee, C.Y., & Wang, C.C. (2008). CPT-based liquefaction assessment by using fuzzy-neural network. Journal of Marine Science and Technology, 16(2), 6. https://doi.org/10.51400/2709-6998.2024.

38. Mustafa, R., Prasad Singh, A., & Ghani, S. (2025). Liquefaction assessment of soil based on SPT and CPT data using novel machine learning techniques: a practical solution. Modeling Earth Systems and Environment, 11(3), 218. https://doi.org/10.1007/s40808-025-02404-0.

39. Kumar, D.R., Samui, P., Burman, A., Wipulanusat, W., Keawsawasvong, S. (2023). Liquefaction susceptibility using machine learning based on SPT data. Intell. Syst. Appl, 20, 200281. https://doi.org/10.1016/j.iswa.2023.200281.

40. Jas, K., Dodagoudar, G.R. (2023b). Explainable machine learning model for liquefaction potential assessment of soils using XGBoost-SHAP. Soil Dyn. Earthq. Eng, 165, 107662. https://doi.org/10.1016/j.soildyn.2022.107662.

41. Katsuumi A., Cong Y., Inazumi S. (2024). AI-Driven Prediction and Mapping of Soil Liquefaction Risks for Enhancing Earthquake Resilience in Smart Cities. Smart Cities, 7(4), 1836-1856. https://doi.org/10.3390/smartcities7040071.

42. Liu C.Y., Ku C.Y., Chiu Y.J., et al. (2024). Evaluation of liquefaction potential in central Taiwan using random forest method. Sci. Rep, 14(1), 27517. https://doi.org/10.1038/s41598-024-79127-2.

43. Chen, M., Kang, X., Ma, X. (2023). Deep Learning–Based Enhancement of Small Sample Liquefaction Data. Int. J. Geomech, 23, 04023140. https://doi.org/10.1061/IJGNAI.GMENG-8381.

44. Bukaita, W., Garcia de Celis, G., & Gurram, M. (2024). Training-Testing Data Ratio Selection for Accurate Time Series Forecasting: A COVID-19 Case Study. In Proceedings of the Future Technologies Conference (pp. 227-246). Cham: Springer Nature Switzerland. https://doi.org/10.1007/978-3-031-73125-9_14.

45. Pham B.T., Qi C., Ho L.S., et al. (2020). A novel hybrid soft computing model using random forest and particle swarm optimization for estimation of undrained shear strength of soil. Sustainability, 12(6), 2218. https://doi.org/10.3390/su12062218.

46. Tien B.D., Hoang N.D., Nhu V.H. (2019). A swarm intelligence-based machine learning approach for predicting soil shear strength for road construction: a case study at Trung Luong National Expressway Project (Vietnam). Eng Comput-germany, 35(3), 955-965. https://doi.org/10.1007/s00366-018-0643-1.

47. Heung B., Ho H.C., Zhang J., et al. (2016). An overview and comparison of machine-learning techniques for classification purposes in digital soil mapping. Geoderma, 265, 62-77. https://doi.org/10.1016/j.geoderma.2015.11.014.

48. Yu, X., Wang, M., Ning, C., & Ji, K. (2025). Predicting largest expected aftershock ground motions using automated machine learning (AutoML)-based scheme, Sci. Rep. 15(1), 942. https://doi.org/10.1038/s41598-024-84668-7.

49. Cetin, K., Der Kiureghian, A., Seed, R.B. (2002). Probabilistic models for the initiation of seismic soil liquefaction. Struct. Saf, 24, 67–82. https://doi.org/10.1016/S0167-4730(02)00036-X.

50. Hu J.L., Tang X.W., Qiu J.N. (2015). A Bayesian network approach for predicting seismic liquefaction based on interpretive structural modeling. Georisk: Assessment and Management of Risk for Engineered Systems and Geohazards, 9(3), 200-217. https://doi.org/10.1080/17499518.2015.1076570.

51. Hu J.L., Huang L.M., Shao, Q. (2025). Combination models of random forest for predicting seismic liquefaction based on SPT, CPT, Vs databases considering sampling strategies. Soil Dynamics and Earthquake Engineering, 198, 109642. https://doi.org/10.1016/j.soildyn.2025.109642.